\documentclass{article}
\usepackage{setspace}
\usepackage[utf8]{inputenc}
\usepackage{amsfonts,amsmath,amssymb,mathtools}
\usepackage[left=2cm,right=2cm,top=2cm,bottom=2cm,bindingoffset=0cm]{geometry}
\usepackage{graphicx}
\usepackage{authblk}
\usepackage{xcolor}
\usepackage{hyperref}
\usepackage{authblk}
\usepackage{slashed}
\usepackage{braket}
\usepackage{comment}
\usepackage{pb-diagram}
\usepackage{wasysym}
\usepackage{amsthm}
\usepackage{cancel}
\usepackage{tikz}\usetikzlibrary{tikzmark}
\usetikzlibrary{fadings} 
\usetikzlibrary{positioning}
\usetikzlibrary{backgrounds} 
\usetikzlibrary{decorations.pathreplacing}
\usetikzlibrary{arrows}
\definecolor{airforceblue}{rgb}{0.36, 0.54, 0.66}	
\definecolor{beige}{rgb}{0.96, 0.96, 0.86}
\definecolor{bittersweet}{rgb}{1.0, 0.44, 0.37}
\definecolor{melon}{rgb}{0.99, 0.74, 0.71}
\definecolor{mustard}{rgb}{1.0, 0.86, 0.35}
\definecolor{lava}{rgb}{0.81, 0.06, 0.13}
\definecolor{magnolia}{rgb}{0.97, 0.96, 1.0}
\definecolor{lavendermist}{rgb}{0.9, 0.9, 0.98}
\definecolor{lavendergray}{rgb}{0.77, 0.76, 0.82}
\definecolor{palepink}{rgb}{0.98, 0.85, 0.87}
\definecolor{palesilver}{rgb}{0.79, 0.75, 0.73}
\definecolor{cadetgrey}{rgb}{0.57, 0.64, 0.69}
\definecolor{anti-flashwhite}{rgb}{0.95, 0.95, 0.96}
\colorlet{Light0anti-flashwhite}{anti-flashwhite!70!white}
\colorlet{Lightanti-flashwhite}{anti-flashwhite!50!white}
\colorlet{Light2anti-flashwhite}{anti-flashwhite!30!white}
\definecolor{linkcolor}{rgb}{0,0,1}
\definecolor{urlcolor}{rgb}{0,0,1}

\usepackage{tikz-cd}
\usetikzlibrary{decorations.markings}
\usetikzlibrary{positioning}
\usepackage{braids}
\usepackage{array}
\usepackage{ytableau}
\usepackage{longtable}
\usepackage{empheq}
\usepackage{multicol}
\usepackage{mathrsfs}
\usepackage{diagbox}
\usepackage[vcentermath]{youngtab}
\usepackage[natbib=true, backend = bibtex, style=numeric-comp, sorting=none]{biblatex}
\hypersetup{pdfstartview=FitH,  linkcolor=linkcolor,urlcolor=urlcolor,citecolor=urlcolor, colorlinks=true}

\newcommand\bem{\begin{pmatrix}}
\newcommand\eem{\end{pmatrix}}
\newcommand\beq{\begin{equation}}
\newcommand\eeq{\end{equation}}
\newcommand\beqs{\begin{equation*}}
\newcommand\eeqs{\end{equation*}}

\date{}

\def\be{\begin{eqnarray}}
\def\ee{\end{eqnarray}}
\def\nn{\nonumber}

\definecolor{red}{rgb}{1,0,0}
\definecolor{orange}{rgb}{1,0.5,0}
\definecolor{violet}{rgb}{0.7,0,1}

\begin{document}

\title{\bf Defect and degree of the Alexander polynomial }
\author[1,2,4]{{\bf E.~Lanina}\thanks{\href{mailto:lanina.en@phystech.edu}{lanina.en@phystech.edu}}}
\author[1,2,3,4]{{\bf A. Morozov}\thanks{\href{mailto:morozov@itep.ru}{morozov@itep.ru}}}

\vspace{5cm}

\affil[1]{Moscow Institute of Physics and Technology, 141700, Dolgoprudny, Russia}
\affil[2]{Institute for Theoretical and Experimental Physics, 117218, Moscow, Russia}
\affil[3]{Institute for Information Transmission Problems, 127994, Moscow, Russia}
\affil[4]{NRC "Kurchatov Institute", 117218, Moscow, Russia}
\renewcommand\Affilfont{\itshape\small}

\maketitle

\vspace{-7cm}

\begin{center}
	\hfill MIPT/TH-08/22\\
	\hfill ITEP/TH-11/22\\
	\hfill IITP/TH-09/22
\end{center}

\vspace{4cm}

\begin{abstract}

{
Defect  characterizes the depth of factorization of terms in differential (cyclotomic) expansions
of knot polynomials, i.e. of the non-perturbative Wilson averages in the Chern-Simons theory.
We prove the conjecture that the defect can be alternatively described as the degree in $q^{\pm 2}$
of the fundamental Alexander polynomial, which formally corresponds to the case of no colors.
We also pose a question if these Alexander polynomials can be arbitrary
integer polynomials of a given degree.
A first attempt to answer the latter question
is a preliminary analysis of antiparallel descendants
of the 2-strand torus knots, which provide a nice set of examples for all values
of the defect. The answer turns out to be positive in the case of defect zero knots, what can be observed already in the case of twist knots. This proved conjecture also allows us to provide a complete set of $C$-polynomials
for the symmetrically colored Alexander polynomials for defect zero.
In this case, we achieve a complete separation of representation and knot variables.
}
\end{abstract}
\setcounter{equation}{0}
\section{Introduction}

The HOMFLY-PT polynomials~\cite{alexander1928topological,leech1970computational,jones1983invent,freyd1985new,kauffman1987state,przytycki1987kobe,morozov2016there,Mironov3_2016,Morozov_2020,arxiv.2205.05650} are the main non-perturbative observables in the Chern-Simons theory \cite{CS,Witten},
and at the moment they are the most important source of information about
such quantities --
much worse understood than correlators in matrix models \cite{UFN3,superintegrarev}
and intimately related supersymmetric (AGT-controlled) low-energy theories \cite{MIRONOV_2010,arxiv.2207.08242}.
Being originally defined as certain contractions of quantum ${\cal R}$-matrices
\cite{reshetikhin1990ribbon,guadagnini1990clausthal,turaev1992state,smirnov2012notes,morozov2010chern,Dunin_Barkowski_2013,mironov2013character,anokhina2013colored,mironov2015colored,mironov2015towards,nawata2017colored}, the HOMFLY polynomials possess a lot of additional structures,
which are not immediately obvious from the Chern-Simons formulation
and are instead the implications of generic representation theory.
They range from the very polynomiality of Wilson loop averages
in appropriate variables $q=\exp\left(\frac{2\pi i}{g+N}\right)$ and $A=q^N$,
to description in terms of Khovanov-Rozansky complexes \cite{khovanov2000categorification,bar2002khovanov,khovanov2004sl,khovanov2007virtual,khovanov2010categorifications,dolotin2014introduction}.
Of special interest in this list is the {\it differential expansion} and its properties.

The differential expansion (DE) for the reduced HOMFLY polynomial in symmetric representations
for the figure-eight knot $4_1$
was introduced in \cite{IMMM}
and was further generalized to other knots \cite{Morozov_2016,Kononov_2016,Morozov_2018,Kameyama_2020,Morozov_2019,morozov2020kntz,BM1,arxiv.2205.12238}.
Originally it was suggested in the form
\be
{\cal H}^{\cal K}_{[r]} \stackrel{?}{=}
\sum_{j=0}^r \frac{[r]!}{[j]![r-j]!}F^{\cal K}_{[j]}(A,q)
\prod_{i=0}^{j-1}\{Aq^{r+i}\}\{Aq^{i-1}\}
\label{DEor}
\ee
with Laurent polynomials $F^{\cal K}_{[j]}(A,q)$.
Here, as usual, $\{x\}:=x-x^{-1}$ and $[n] := \{q^n\}/\{q\}$.
The DE is also known under the name of {\it cyclotomic expansion} \cite{garoufalidis2005analytic,garoufalidis2011asymptotics,arxiv.1512.07906,Kameyama_2020,arxiv.1908.04415,arxiv.2101.08243,arxiv.2110.03616}.
The question mark in (\ref{DEor}) stands because the relation is not quite true
for all knots ${\cal K}$, in generic case it is actually weaker \cite{Kononov_2015}:
\be
{\cal H}^{\cal K}_{[r]} =
\sum_{j=0}^r \frac{[r]!}{[j]![r-j]!}G^{\cal K}_{[j]}(A,q)
 \{A/q\}\prod_{i=0}^{j-1}\{Aq^{r+i}\}\,.
\label{DE}
\ee
This DE for symmetric representations actually follows from representation theory,
and it involves polynomial $G^{\cal K}_{[j]}$ rather than $F^{\cal K}_{[j]}$.
The defect \cite{Kononov_2015} characterizes the degree of its enhancement towards $F^{\cal K}_{[j]}$.

The variables $G^{\cal K}_{[j]}$ can provide better coordinates in the space of knots
than the HOMFLY polynomials themselves \cite{artdiff},
still they are far from being {\it free} parameters.
In particular, they satisfy the $C$-polynomial equations \cite{arxiv.2205.12238,Garoufalidis_2006,Cpols},
which are still far from being well understood and classified.

In this paper, we clarify the situation with an alternative description of the defect $\delta$ \cite{Kononov_2015} --
as dictated by the degree $\delta+1$ of the fundamental Alexander polynomial in $q^{\pm 2}$.
We explicitly prove that this is indeed a corollary of definition (\ref{DE})
and derive explicit expressions for the DE coefficients $G^{\mathcal{K}}_{[r]}(1,q)$ at $A=1$
in terms of the coefficients of the fundamental Alexander polynomial.
This description provides an explicit restriction on $G^{\mathcal{K}}_{[r]}(1,q)$ -- they are not free.
What still {\it can} be free are the fundamental coefficients $G_{[1]}(1,q)=F_{[1]}(1,q)$\footnote{We sometimes omit the upper index $\mathcal{K}$ to shorten expressions.}
at $r=1$ --
they could be {\it arbitrary} integer symmetric polynomials of degree $\delta$.
We show that this is indeed the case for $\delta=0$ and provide some ideas on how this problem can be investigated for higher $\delta$.
Namely, we use the defect-preserving antiparallel evolution \cite{2204.05977} of the
2-strand knots, which provides rich sets of examples for all values of the defect.
This also allows us to highlight the properties of the antiparallel evolution
and derive non-trivial  examples, when the defect drops down at particular points
(the conjecture of \cite{2204.05977} does not allow it to {\it raise}
but allows occasional dropdowns).

The paper is organized as follows. In Sections \ref{orDE}, \ref{defdef}
we remind the definitions of the DE and its defect for the case of
symmetric representations. The central Section \ref{defAp} proves that defect $\delta$ is unambiguously related
to the degree $\delta+1$ of the fundamental Alexander , and in Subection \ref{exprforG}
we express all the coefficients $G_{[r]}(1,q)$
of DE for the symmetric Alexander polynomials through $G_{[1]}(1,q)$.
These are our main results.

The remaining part of the paper concerns open questions and is not yet conclusive.
We discuss the following issues.

$\bullet$ First, if all $G_{[r]}(1,q)$ are expressed through $G_{[1]}(1,q)$, can the latter
one be arbitrary (free)? It is for $\delta=0$, but what about the higher defects? To address this question, in Section \ref{integra} we rewrite
the expressions from Subection \ref{exprforG} in a more explicit/detailed form, and
consider a similarly rich family of antiparallel descendants
of the two-strand torus knots.
This is an interesting family by itself, but the corresponding sets
$G_{[1]}(1,q)$ for $\delta \geq 1$  look unexpectedly restricted.
At the moment, it is unclear if this is a property of this particular set
(it is not as big as it seems) or  $G_{[1]}(1,q)$ are indeed far from being free.

$\bullet$ Second, is the defect indeed as simple as described in Section \ref{defdef}, or is there
a more sophisticated structure, associated with more complicated degeneracy diagrams?
We do not go deep into this problem (raised already in \cite{Kononov_2015} and \cite{2204.05977}),
we just provide some illustrations of "accidental" degeneracies in Section \ref{defpre}. This also helps to reformulate the statement about
invariance of the defect under antiparallel evolution more carefully -- 
defect can actually drop down
at some evolution points, and there remains a question, if this should be treated
as "accidents" or as a manifestation of additional structures,
promoting defect from just a number to a slightly more involved characteristic of the DE.

$\bullet$ Third, can we express the fact that all $G_{[r]}(1,q)$ are made from $G_{[1]}(1,q)$
as some equations for $G_{[r]}(1,q)$?
In Section \ref{Cpols}, we show that such "Alexander $C$-polynomials" can indeed be defined
and they are considerably simpler than generic $C$-polynomials for the HOMFLY polynomial and
superpolynomials.

$\bullet$ Fourth, can we lift the description of $G_{[r]}(1,q)$ through  $G_{[1]}(1,q)$
to non-symmetric representations?
In Section \ref{otherreps}, we make first steps in this direction.

Conclusion in Section \ref{conc} contains a brief summary.

\setcounter{equation}{0}
\section{On the origins of the differential expansion
\label{orDE}}

In this paper, we mostly consider the differential expansion of the HOMFLY polynomials
in the simplest case of symmetric representations.
All the HOMFLY polynomials are \textit{reduced}, i.e. they are Wilson averages in an irreducible representation $R$,
divided by the dimension of the representation. The differential (cyclotomic) expansion
is what follows from elementary representation theory and from the fact that it is respected by the HOMFLY polynomials. Namely, we get use of the properties below.
\begin{itemize}
    \item In $\mathfrak{sl}_N$ algebra, $[R_1,\dots,R_N]=[R_1+\delta R,\dots,R_N+\delta R]$ for any integer $\delta R\,$.
    \item In $\mathfrak{sl}_N$ algebra, a representation $R$ and its conjugate $\bar{R}$ are equivalent.
    \item The HOMFLY polynomial possesses the symmetry under the transposition of a diagram~\cite{arxiv.1012.2636}:
    \be\label{transposition}
\mathcal{H}^{\mathcal{K}}_{R^\vee}(q,A) = \mathcal{H}^{\mathcal{K}}_R(q^{-1},A)\,,
\ee
where $R^\vee$ denotes the transposition of a Young diagram $R$.
\end{itemize}
These facts restrict the HOMFLY group structure drastically. We use them to obtain the differential expansion for the symmetric HOMFLY polynomials. To begin with, for $N=1$ all representations are trivial and the HOMFLY polynomials are unities.
In particular, for $R=[r]$ we have
\be
\mathcal{H}_{[r]}^{\mathcal{K}}(q, A=q)=1 \;\Longrightarrow \; \mathcal{H}_{[r]}^{\mathcal{K}}(q, A)-1 \;\vdots\;\{A / q\}\,.
\label{Hr1}
\ee
At the same time, for $N=r$ the antisymmetric representation $[1^r]$ is the same as singlet, $\emptyset$, so
\be
\mathcal{H}_{[1^r]}^{\mathcal{K}}(q, A=q^r)=1 \; \Longrightarrow \; \mathcal{H}_{[1^r]}^{\mathcal{K}}(q, A)-1 \;\vdots\;\{A / q^r\}\,,
\label{Hr2}
\ee
and utilizing the basic transposition property~\eqref{transposition} and
taking into account (\ref{Hr1}), we obtain:
\be \label{Hr}
\mathcal{H}_{[r]}^{\mathcal{K}}(q, A)-1\sim \{A q^r\}\{A / q\}\,.
\ee
In fact, for $r>1$ this is not the full story.
We illustrate this by the next simplest example.
For $N=3$, we have  an additional property:
$\mathcal{H}_{[1]}^{\mathcal{K}}(q, A=q^3)=\mathcal{H}_{[1,1]}^{\mathcal{K}}(q, A=q^3)$,
because in this case the two representations $[1]$ and $[1,1]$ are the same.
This fact implies that
\begin{equation}
\mathcal{H}_{[1,1]}-\mathcal{H}_{[1]} \sim\left\{A / q^{3}\right\} \; \stackrel{\eqref{transposition}}{\Longleftrightarrow}\; \mathcal{H}_{[2]}-\mathcal{H}_{[1]} \sim\left\{A q^{3}\right\} \stackrel{\eqref{Hr}}{\Longrightarrow} \mathcal{H}_{[2]}-\mathcal{H}_{[1]} \sim\left\{A q^{3}\right\}\{A / q\}\,.
\end{equation}
From~\eqref{Hr}, we get
\begin{equation}
	\mathcal{H}_{[1]}^{\mathcal{K}}(q, A)=1+\{A q\}\{A / q\}F_{[1]}^{\mathcal{K}}(q, A)\,.
\end{equation}
with Laurent polynomial $F_{[1]}(q,A)$.
Then
\begin{equation}
	\mathcal{H}_{[2]}^{\mathcal{K}}(q, A)=\mathcal{H}_{[1]}^{\mathcal{K}}(q, A)+g_2^{\mathcal{K}}(q, A)\{A q^{3}\}\{A / q\}=1+\Big(\{A  q\}F_{[1]}^{\mathcal{K}}(q, A)+h_2^{\mathcal{K}}(q, A)\{A q^{3}\}\Big)\{A / q\}\,.
\end{equation}
From~\eqref{Hr}, it follows that $\mathcal{H}_{[2]}^{\mathcal{K}}(q, A)-1\sim \{A q^2\}\{A / q\}$,
so we set $h_{2}=F_{[1]}+\left\{A q^{2}\right\}G_{[2]}$
with some polynomial $G_{[2]}(q,A)$ and obtain
\begin{equation}
\begin{array}{r}
\mathcal{H}_{[1]}=1+F_{[1]} \cdot\{A q\}\{A / q\}\,, \\ \\
\mathcal{H}_{[2]}=1+[2] F_{[1]} \cdot\left\{A q^{2}\right\}\{A / q\}
+G_{[2]} \cdot\left\{A q^{3}\right\}\left\{A q^{2}\right\}\{A / q\}\,.
\end{array}
\end{equation}
In the same way, one can iteratively deduce that for an arbitrary knot
\begin{equation}\label{DEHr}
\boxed{\mathcal{H}_{[r]}^{\mathcal{K}}=\sum_{k=0}^{r} \frac{[r] !}{[k] ![r-k] !} \cdot G_{[k]}^{\mathcal{K}}(A, q) \cdot\left(\prod_{i=0}^{k-1}\left\{A q^{r+i}\right\}\right)\{A / q\}}
\end{equation}
with some polynomial factors $G_{[k]}^{\mathcal{K}}(A, q)\,$.

\setcounter{equation}{0}
\section{Defect of the differential expansion of the symmetric HOMFLY
\label{defdef}}

Now, there comes something else -- perhaps, a little closer to the mysterious nature of {\it knots}.
The coefficients $G_{[k]}$ turn out to factorize further.
Their factorization property depends on the knot parameter $\delta^\mathcal{K}$ named \textit{defect}:
\be
\boxed{
G_{[s]}^{\mathcal{K}^{(\delta)}}(A, q)={\cal G}_{[s]}^{\mathcal{K}^{(\delta)}}(A, q) \cdot
\prod_{i=1}^{{\rm floor}\left(\frac{s-1}{\delta+1}\right)}\left\{A q^{i-1}\right\}\,.
}
\label{Gdelta}
\ee
Factorization is maximal for $\delta=0$, the corresponding coefficients\footnote{To shorten the notations, we sometimes use $(\delta)$ instead of $\mathcal{K}^{(\delta)}$ superscript.} ${\cal G}^{(0)}_{[s]}$ are usually
denoted by $F_{[s]}$.
For $s=1$ always $G_{[1]}=F_{[1]}$, we use this fact to distinguish between the first and
all other coefficients of the differential expansion.

The property (\ref{Gdelta}) can be depicted by the following ladder diagrams:
\newpage
\be\label{d=0}
\begin{picture}(300,70)(-40,-30)
\put(0,-20){
\put(-90,0){\mbox{defect $\delta^{\cal K}=0$:}}
\put(-100,0){
\put(177,-13){\circle*{5}}
\put(190,-20){\line(1,0){110}}
\put(190,-5){\line(1,0){110}}
\put(215,10){\line(1,0){85}}
\put(240,25){\line(1,0){60}}
\put(265,40){\line(1,0){35}}
\put(290,55){\line(1,0){10}}
%
\put(190,-20){\line(0,1){15}}
\put(215,-20){\line(0,1){30}}
\put(240,-20){\line(0,1){45}}
\put(265,-20){\line(0,1){60}}
\put(290,-20){\line(0,1){75}}
%
\put(194,-16){\mbox{\footnotesize $\{A\}$}}
\put(219,-16){\mbox{\footnotesize $\{A\}$}}
\put(244,-16){\mbox{\footnotesize $\{A\}$}}
\put(269,-16){\mbox{\footnotesize $\{A\}$}}
\put(218,-1){\mbox{\footnotesize $\{Aq\}$}}
\put(243,-1){\mbox{\footnotesize $\{Aq\}$}}
\put(268,-1){\mbox{\footnotesize $\{Aq\}$}}
\put(241,14){\mbox{\footnotesize $\{Aq^2\}$}}
\put(266,14){\mbox{\footnotesize $\{Aq^2\}$}}
\put(266,29){\mbox{\footnotesize $\{Aq^3\}$}}
%
\put(145,-35){\mbox{$s$}}
\put(175,-35){\mbox{{\footnotesize $1$}}}
\put(200,-35){\mbox{{\footnotesize $2$}}}
\put(225,-35){\mbox{{\footnotesize $3$}}}
\put(250,-35){\mbox{{\footnotesize $4$}}}
\put(275,-35){\mbox{{\footnotesize $5$}}}
\put(300,-35){\mbox{{\footnotesize $6$}}}
%
\put(320,-15){\mbox{{\footnotesize $1$}}}
\put(320,0){\mbox{{\footnotesize $2$}}}
\put(320,15){\mbox{{\footnotesize $3$}}}
\put(320,30){\mbox{{\footnotesize $4$}}}
\put(320,45){\mbox{{\footnotesize $5$}}}
%
%
}}
\end{picture}
\ee

\be\label{d=1}
\begin{picture}(300,70)(-40,-20)
\put(0,-20){
\put(-90,0){\mbox{defect $\delta^{\cal K}=1$:}}
\put(-100,0){
\put(194,-16){\circle*{3}}
\put(185,-16){\circle*{5}}
\put(200,-20){\line(1,0){115}}
\put(200,-10){\line(1,0){115}}
\put(220,0){\line(1,0){95}}
\put(240,10){\line(1,0){75}}
\put(260,20){\line(1,0){55}}
\put(280,30){\line(1,0){35}}
\put(300,40){\line(1,0){15}}
\put(200,-20){\line(0,1){10}}
\put(210,-20){\line(0,1){10}}
\put(220,-20){\line(0,1){20}}
\put(230,-20){\line(0,1){20}}
\put(240,-20){\line(0,1){30}}
\put(250,-20){\line(0,1){30}}
\put(260,-20){\line(0,1){40}}
\put(270,-20){\line(0,1){40}}
\put(280,-20){\line(0,1){50}}
\put(290,-20){\line(0,1){50}}
\put(300,-20){\line(0,1){60}}
\put(310,-20){\line(0,1){60}}
%
\put(160,-35){\mbox{$s$}}
\put(183,-35){\mbox{{\footnotesize $1$}}}
\put(193,-35){\mbox{{\footnotesize $2$}}}
\put(203,-35){\mbox{{\footnotesize $3$}}}
\put(213,-35){\mbox{{\footnotesize $4$}}}
\put(223,-35){\mbox{{\footnotesize $5$}}}
\put(233,-35){\mbox{{\footnotesize $6$}}}
\put(243,-35){\mbox{{\footnotesize $7$}}}
\put(253,-35){\mbox{{\footnotesize $8$}}}
\put(263,-35){\mbox{{\footnotesize $9$}}}
\put(270,-35){\mbox{{\footnotesize $10$}}}
\put(280,-35){\mbox{{\footnotesize $11$}}}
\put(290,-35){\mbox{{\footnotesize $12$}}}
\put(300,-35){\mbox{{\footnotesize $13$}}}
\put(310,-35){\mbox{{\footnotesize $14$}}}
\put(320,-18){\mbox{{\footnotesize $1$}}}
\put(320,-8){\mbox{{\footnotesize $2$}}}
\put(320,2){\mbox{{\footnotesize $3$}}}
\put(320,12){\mbox{{\footnotesize $4$}}}
\put(320,22){\mbox{{\footnotesize $5$}}}
\put(320,32){\mbox{{\footnotesize $6$}}}
%
%
}}
\end{picture}
\ee

\be\label{d=2}
\begin{picture}(300,80)(-40,0)
\put(0,-20){
\put(-90,20){\mbox{defect $\delta^{\cal K}=2$:}}
\put(-100,30){
\put(204,-16){\circle*{3}}
\put(195,-16){\circle*{3}}
\put(185,-16){\circle*{5}}
\put(210,-20){\line(1,0){115}}
\put(210,-10){\line(1,0){115}}
\put(240,0){\line(1,0){85}}
\put(270,10){\line(1,0){55}}
\put(300,20){\line(1,0){25}}
%
\put(210,-20){\line(0,1){10}}
\put(220,-20){\line(0,1){10}}
\put(230,-20){\line(0,1){10}}
\put(240,-20){\line(0,1){20}}
\put(250,-20){\line(0,1){20}}
\put(260,-20){\line(0,1){20}}
\put(270,-20){\line(0,1){30}}
\put(280,-20){\line(0,1){30}}
\put(290,-20){\line(0,1){30}}
\put(300,-20){\line(0,1){40}}
\put(310,-20){\line(0,1){40}}
\put(320,-20){\line(0,1){40}}
\put(160,-35){\mbox{$s$}}
\put(183,-35){\mbox{{\footnotesize $1$}}}
\put(193,-35){\mbox{{\footnotesize $2$}}}
\put(203,-35){\mbox{{\footnotesize $3$}}}
\put(213,-35){\mbox{{\footnotesize $4$}}}
\put(223,-35){\mbox{{\footnotesize $5$}}}
\put(233,-35){\mbox{{\footnotesize $6$}}}
\put(243,-35){\mbox{{\footnotesize $7$}}}
\put(253,-35){\mbox{{\footnotesize $8$}}}
\put(263,-35){\mbox{{\footnotesize $9$}}}
\put(270,-35){\mbox{{\footnotesize $10$}}}
\put(280,-35){\mbox{{\footnotesize $11$}}}
\put(290,-35){\mbox{{\footnotesize $12$}}}
\put(300,-35){\mbox{{\footnotesize $13$}}}
\put(310,-35){\mbox{{\footnotesize $14$}}}
\put(330,-18){\mbox{{\footnotesize $1$}}}
\put(330,-8){\mbox{{\footnotesize $2$}}}
\put(330,2){\mbox{{\footnotesize $3$}}}
\put(330,12){\mbox{{\footnotesize $4$}}}
%
%
}}
\end{picture}
\ee

\be\label{d=3}
\begin{picture}(300,70)(-40,20)
\put(0,-20){
\put(-90,40){\mbox{defect $\delta^{\cal K}=3$:}}
\put(-100,50){
\put(214,-16){\circle*{3}}
\put(205,-16){\circle*{3}}
\put(195,-16){\circle*{3}}
\put(185,-16){\circle*{5}}
\put(220,-20){\line(1,0){105}}
\put(220,-10){\line(1,0){105}}
\put(260,0){\line(1,0){65}}
\put(300,10){\line(1,0){25}}
%
\put(220,-20){\line(0,1){10}}
\put(230,-20){\line(0,1){10}}
\put(240,-20){\line(0,1){10}}
\put(250,-20){\line(0,1){10}}
\put(260,-20){\line(0,1){20}}
\put(270,-20){\line(0,1){20}}
\put(280,-20){\line(0,1){20}}
\put(290,-20){\line(0,1){20}}
\put(300,-20){\line(0,1){30}}
\put(310,-20){\line(0,1){30}}
\put(320,-20){\line(0,1){30}}
\put(160,-35){\mbox{$s$}}
\put(183,-35){\mbox{{\footnotesize $1$}}}
\put(193,-35){\mbox{{\footnotesize $2$}}}
\put(203,-35){\mbox{{\footnotesize $3$}}}
\put(213,-35){\mbox{{\footnotesize $4$}}}
\put(223,-35){\mbox{{\footnotesize $5$}}}
\put(233,-35){\mbox{{\footnotesize $6$}}}
\put(243,-35){\mbox{{\footnotesize $7$}}}
\put(253,-35){\mbox{{\footnotesize $8$}}}
\put(263,-35){\mbox{{\footnotesize $9$}}}
\put(270,-35){\mbox{{\footnotesize $10$}}}
\put(280,-35){\mbox{{\footnotesize $11$}}}
\put(290,-35){\mbox{{\footnotesize $12$}}}
\put(300,-35){\mbox{{\footnotesize $13$}}}
\put(310,-35){\mbox{{\footnotesize $14$}}}
\put(330,-18){\mbox{{\footnotesize $1$}}}
\put(330,-8){\mbox{{\footnotesize $2$}}}
\put(330,2){\mbox{{\footnotesize $3$}}}
%
%
}}
\end{picture}
\ee

\be
\ldots
\nn
\ee

It looks like a highly non-trivial structure,
and neither its origins nor its universality are understood.
Also unknown is identification of the defect with more conventional
discrete topological numbers, which are associated with knots. 

As we will see in Section \ref{defpre}, the list of diagrams above
can be \textit{not exhaustive}: sometimes some boxes can be eliminated --
there are additional zeroes in coefficients of the DE.
Moreover, there can be some regularity -- new zeroes appear at the
boundaries of our diagrams, and preserve their ladder structure.
This can be a signal that defect is not just a number, but a more complicated
structure, perhaps, like a Young diagram, as suggested in original \cite{Kononov_2015}. These additional zeroes also affect precise formulation of
antiparallel invariance of the defect\footnote{In~\cite{2204.05977}, it was conjectured that the defect does not change under substitution of any knot crossing with an antiparallel braid, except for a few knots for which the defect occasionally drops down.}, which is also discussed in Section \ref{defpre}.

\bigskip

In this paper, we address/prove still another -- and simpler, characterization of the defect:
the conjecture \cite{Kononov_2015} that it is equal to the degree of
the fundamental Alexander polynomial minus one.
This is a relatively simple statement, it concerns only the bottom level of the above diagrams.
Indeed, the Alexander polynomial arises at $A=1$ and then (\ref{Gdelta})
means that there are just a few non-vanishing
\be
G_{[s]}^{\mathcal{K}^{(\delta)}}(1, q)={\cal G}_{[s]}^{\mathcal{K}^{(\delta)}}(1, q)\,,
\ \ \ \ \ \ \ 1<s\leq \delta+1
\ee
shown by small black dots in pictures~\eqref{d=0}-\eqref{d=3}.
Big dots stand for $F_{[1]}(1,q)$, which can vanish only {\it occasionally}
(see Section \ref{defpre} below).
For higher $s$, the product at the r.h.s. of (\ref{Gdelta}) contains $\{A\}$, which
vanishes at $A=1$.
Despite the fact that the statement seems simple,
it becomes highly informative
if one assumes that (\ref{Gdelta}) is known --
then it allows to predict a lot about the colored HOMFLY polynomial
(and, actually, superpolynomials as well)
by making a nearly trivial calculation for the fundamental Alexander polynomial.

\setcounter{equation}{0}
\section{Defect defines the degree of the fundamental Alexander polynomial}\label{defAp} 
In this section, we prove that the defect (in most cases, see Section~\ref{defpre}) can be defined as the degree of the fundamental Alexander polynomial minus one. It turns out that it is a consequence of two facts.
\begin{itemize}
    \item The reduced symmetric\footnote{This symmetry actually holds for all single-hook representations.} Alexander polynomial possesses the following property:
\be
\mathcal{A}_{[r]}(q) = \mathcal{A}_{[1]}(q^r)\,.
\label{symAl}
\ee
\item The DE~\eqref{DEHr} for the symmetric Alexander polynomial ($A=1$) becomes
\begin{equation}\label{AlDE}
    \mathcal{A}_{[r]}^{\mathcal{K}}=\sum_{k=0}^{\min(r,\delta+1)} \frac{[r] !}{[k] ![r-k] !} \cdot G_{[k]}^{\mathcal{K}}(1, q) \cdot\left(\prod_{i=0}^{k-1}\left\{q^{r+i}\right\}\right)\{q^{-1}\}
\end{equation}
with $F_{[1]}(1,q)=G_{[1]}(1,q)$. In particular
\be\label{Al1}
\mathcal{A}_{[1]} = 1 - F_{[1]}(1,q)\{q\}^2\,,
\label{HfundDE}
\ee
where recall that $\{x\} := x-x^{-1}$, so that the quantum numbers are $[n]:=\frac{\{q^n\}}{\{q\}}$.
\end{itemize}

\subsection{Small defect examples}
Before providing a proof for the general defect $\delta$, we parse examples for small values of the defect in order to clarify the arguments.

\medskip

$\bullet$ {\bf For the defect $\delta=0$} the DE~\eqref{AlDE} contains two terms for
all symmetric representations $[r]$:
\be\label{Ald=0}
\mathcal{A}_{[r]} = 1 - [r] F_{[1]}(1,q)\{q^r\}\{q\}\,.
\ee
Then (\ref{symAl}), \eqref{Al1}, \eqref{Ald=0} imply
\be
[r]\{q^r\}\{q\} F_{[1]}(1,q) = \{q^r\}^2 F_{[1]}(1,q^r)
\ \ \ \Longrightarrow \ \ \
F_{[1]}(1,q^r) = F_{[1]}(1,q) = {\rm const} =a\,.
\label{def0a}
\ee
The trick is to look at the $r$ dependence, which at the l.h.s. is just quadratic in $q^{\pm r}$
because $F_{[1]}(1,q) $ is independent of $r$.
This restricts the $q$-dependence of $F_{[1]}(1,q^r)$ at the r.h.s --
actually to nothing.

Note that from~\eqref{def0a}, it follows that the \textbf{degree of the fundamental Alexander polynomial} is $\delta+1=1$. However, sometimes it happens that $F_{[1]}(1,q)=0$ and the degree of the fundamental Alexander polynomial drops to $0$. Examples are provided in Subsection~\ref{3_1Des}.

\medskip

$\bullet$ {\bf For the defect $\delta=1$} it is released, but just a little.
For $r>1$ there are three terms:
\be
\mathcal{A}_{[r]} = 1 - [r] F_{[1]}(1,q)\{q^r\}\{q\} - \frac{[r][r-1]}{[2]}G_{[2]}(1,q)\{q^{r+1}\}\{q^r\}\{q\}\,,
\ee
and therefore
\be
[r]\{q^r\}\{q\} F_{[1]}(1,q)+ \frac{[r][r-1]}{[2]}G_{[2]}(1,q)\{q^{r+1}\}\{q^r\}\{q\}
= \{q^r\}^2 F_{[1]}(1,q^r)  \ \ \ \Longrightarrow
\nn \\
\ \ \ \Longrightarrow \ \ \
F_{[1]}(1,q) =   b\cdot(q^2+q^{-2})+a, \ \ \ \ \
G_{[2]}(1,q) = b\cdot \{q^2\}\,
\ee
with some constants $a$ and $b$ \footnote{Of course, the $q$-dependence of entire $F_{[1]}(A,q)$ and $G_{[2]}(A,q)$ with $A\neq 1$ can be far more involved.
For example, for the defect-1 knot $[7,5]$:
\be\nonumber
	\begin{aligned}
F_{[1]}^{[7,5]}(A,q) &= -\frac{(A^2 + 1)(A^2q^2 + q^4 - q^2 + 1)}{A^6q^2}\,, \\
		G_{[2]}^{[7,5]}(A,q)\!\! &= \! A^{-13}q^{-14}\{A^{10}q^{12} + A^8(q^{14}+  q^{10} + q^8)+ A^6(q^8 + q^4)
		- A^4(q^{16}+q^{10} +q^4 -q^2)
		- \\
		&-A^2(q^{16} +q^{14}  -2 q^{12} +q^{10} + 3 q^8 -2q^6   +2q^2 -1)
		- (q^{12} - q^{10} - q^8+ 2q^6- q^2 + 1)\}
	\end{aligned}
	\ee
are already quite complicated,
still
\be\nonumber
\begin{aligned}
F_{[1]}^{[7,5]}(1,q) &= -2(q^2+q^{-2})\,, \\
G_{[2]}^{[7,5]}(1,q) &= -2(q^2-q^{-2})= -2\{q^2\}\,. 
\end{aligned}
\ee}. At the l.h.s., we have a quartic polynomial of $q^{\pm r}$, therefore the same must happen to the r.h.s.,
thus $F_{[1]}$ is at most quadratic.
Note that this does not mean, that $G_{[2]}(1,q)$ is independent of $q$ --
the dependence exists, but it is fully fixed in terms of the $q$-dependence of $F_{[1]}(1,q)$.

The conjecture about the defect still holds: the \textbf{degree of the Alexander polynomial} turns out to be $\delta+1=2$.

We return to defect $0$ if $b=0$.
To understand if the conjecture about defect-Alexander relation still holds at such special points,
 it is necessary to check if factorization of entire HOMFLY also increases.
Some examples are provided in Subsection~\ref{5_1Des}.

\medskip

$\bullet$ {\bf For the defect $\delta=2$} and $r>2$ there are four terms:
\be
\mathcal{A}_{[r]}(q) = 1 - [r] F_{[1]}(1,q)\{q^r\}\{q\} - \frac{[r][r-1]}{[2]}G_{[2]}(1,q)\{q^{r+1}\}\{q^r\}\{q\}
- \nn \\
 - \frac{[r][r-1][r-2]}{[2][3]}G_{[3]}(1,q)\{q^{r+2}\}\{q^{r+1}\}\{q^r\}\{q\}
 = {\mathcal A}_{[1]}(q^r) = 1 -  F_{[1]}(1,q^r)\{q^r\}^2\,,
\ee
that implies
\be
 F_{[1]}(1,q^r) - F_{[1]}(1,q) = \{q^{r+1}\}\{q^{r-1}\}\frac{G_{[2]}(1,q)}{\{q^2\}}
 + \{q^{r+2}\}\{q^{r+1}\}\{q^{r-1}\}\{q^{r-2}\}\frac{G_{[3]}(1,q)}{\{q^2\}\{q^3\}}\,,
\ee
and
\be
F_{[1]}(q) = c\cdot(q^4+q^{-4})+b\cdot(q^2+q^{-2})+a\,,
\nn \\
G_{[2]}(q) = \Big(c\cdot(q^4+q^2+q^{-2}+q^{-4}) + b\Big)\cdot\{q^2\}\,,
\nn \\
G_{[3]}(q) = c\cdot\{q^2\}\{q^3\}\,
\ee
with some knot-dependent integers $a$, $b$ and $c$. The \textbf{degree of the Alexander polynomial} is still $\delta+1=3$, except some points, where $c$, $b$, $a$ turn to zero.

For example, for the defect-2 torus knot $7_1$
\be
F_{[1]}^{7_1}(1,q) = -(q^4+q^{-4}) -(q^2+q^{-2})-2\,,
\nn \\
G_{[2]}^{7_1}(1,q) = -(q^4+q^2+q^{-2}+q^{-4})-1\,,
\nn \\
G_{[3]}^{7_1}(1,q) = -1\,,
\ee
i.e. $a=b=c=-1$.

\subsection{The case of generic defect
\label{gendef}}

In this subsection, we raise the examples from the previous subsection to generic $\delta$. Again, our goal is to find the restriction on $F_{[1]}(1,q)$ --
and then it allows to express all non-vanishing $G_{[r]}(1,q)$
through this $F_{[1]}(1,q)$.
Expectedly or not, this expression is rather involved --
it is the usual amusement in knot theory to observe how
fast the complicated structures emerge  from the trivial inputs.
The lessons for a generic quantum field theory are still to be drawn.

\subsubsection{Degree of the Alexander polynomial}\label{DofAl}

Substituting the DE at $A=1$~\eqref{AlDE} into (\ref{symAl}), we get:
\be
 1-[r] F_{[1]}(1,q)\{q^r\}\{q\} -
\sum_{j=2}^{{\rm min}(r,\delta+1)} \frac{[r]!}{[j]![r-j]!} G_{[j]}(1,q)\{q\}\prod_{i=1}^{j}\{q^{r+i-1}\}
= 1-F_{[1]}(1,q^r)\{q^r\}^2\,.
\label{gendeltasymAl}
\ee
This equality must be true for all $r$.
The l.h.s. depends on $q^r$ only through binomial coefficients and differentials,
and it is clear that it is a
polynomial of degree $2j_{\rm max} = 2\delta+2$ in $q^{\pm r}$.
The r.h.s. depends on $q^{\pm r}$ through the factor $\{q^r\}^2$ of degree $2$
and $F_{[1]}(1,q^r)$. Since all the dependencies are actually on even powers of $q$,
this implies that $F_{[1]}(1,q)$ is actually a polynomial of $q^{\pm 2}$ of degree $\delta$, and the whole \textbf{fundamental Alexander polynomial has degree} $\delta+1$. Moreover, since $\mathcal{A}_{[1]}(q)=\mathcal{A}_{[1]}(q^{-1})$~\eqref{transposition}, $F_{[1]}(1,q)$ must be symmetric under the change $q\rightarrow q^{-1}$.

In other words,
\be
F^{(\delta)}_{[1]}(1,q) = a^{(\delta)}_{0} + \sum_{j=1}^\delta a^{(\delta)}_j \cdot (q^{2j}+q^{-2j})
\label{fundAl}
\ee
with $q$-independent integers $a^{(\delta)}_k$.
We also put the label $(\delta)$ on $F_{[1]}(1,q)$ and in what follows -- on $G_{[k]}(1,q)$,
to emphasize that at $A=1$ they depend on the knot (a point in a huge variety of objects)
only through its defect and {\bf just a few additional parameters} $a^{(\delta)}_i$.

Now, the conjecture about the degree of the fundamental Alexander polynomial is proved, and in the next subsection we present explicit expressions of $G_{[k]}(1,q)$ through the coefficients $a^{(\delta)}_i$ of $F_{[1]}(1,q)$.

\subsubsection{Expressions for non-vanishing $G_{[k]}(1,q)$
\label{exprforG}}

Implying~\eqref{fundAl}, we can rewrite (\ref{gendeltasymAl}) as
\be\label{EqG}
 \sum_{j=1}^\delta a^{(\delta)}_j \cdot (q^{2jr}-q^{2j}-q^{-2j}+q^{-2jr})
= \sum_{j=2}^{{\rm min}(r,\delta+1)}  g^{(\delta)}_j(q) \cdot \prod_{i=1}^{j-1}\{q^{r+i}\}\{q^{r-i}\}\,,
\ee
where we denote
\be
g_j^{(\delta)} (q) := \frac{G^{(\delta)}_{[j]}(1,q)}{\prod_{i=2}^j\{q^i\}}\,,\quad 2\leq j\leq \delta+1\,.
\label{defg}
\ee
Equation~\eqref{EqG} must hold for any $r$, so we can extract coefficients at certain powers of $q^r$ and obtain the following system of equations:
\be
	a_\delta^{(\delta)}=g^{(\delta)}_{\delta+1}\,,\nn \\
	a_{\delta-1}^{(\delta)}=g_\delta^{(\delta)}-g_{\delta+1}^{(\delta)}(q^{\delta+1}+q^{-\delta-1})[\delta]\,,\nn \\
	a_{\delta-2}^{(\delta)}=g_{\delta-1}^{(\delta)}-g_{\delta}^{(\delta)}[\delta-1](q^{\delta}+q^{-\delta})
+g_{\delta+1}^\delta\left([\delta]^2+[\delta][\delta-1]\frac{q^{2\delta+2}+q^{-2\delta-2}}{q+q^{-1}}\right),  \nn \\
	\dots
\ee
In general
\be\label{C-pol1}
a_j^{(\delta)}=\sum_{k=0}^{\delta-j}g_{j+k+1}^{(\delta)}\sigma_{j,k}\,,
\ee
where
\be
\sigma_{j,k}
= \sum\limits_{i=0}^k(-1)^k q^{k(k-2i+1)-2i}\frac{(q^{-2};q^{-2})_{j+k}}{(q^{-2};q^{-2})_{i}(q^{-2};q^{-2})_{j+k-i}}
\frac{(q^2;q^2)_{j+k}}{(q^2;q^2)_{k-i}(q^2;q^2)_{j+i}}\,.
\label{sigmaf}
\ee
Note that relations~\eqref{C-pol1} connect knot-factors $F_{[1]}(1,q)$ and $G_{[k]}(1,q)$, so that actually they are the symmetric Alexander $C$-polynomials (see Section~\ref{Cpols}).

One can revert relations~\eqref{C-pol1} and express all $g_j$ through $a_j$:
\be
g^{(\delta)}_{\delta+1} = a^{(\delta)}_\delta\,,
\nn \\
g^{(\delta)}_{\delta} = a^{(\delta)}_{\delta-1}
+ a^{(\delta)}_\delta  \cdot(q^{\delta+1}+q^{-\delta-1})[\delta]\,,
\nn \\
g^{(\delta)}_{\delta-1} = a^{(\delta)}_{\delta-2}
+ a^{(\delta)}_{\delta-1}\cdot  (q^{\delta}+q^{-\delta})[\delta-1]
+ a^{(\delta)}_\delta\cdot \Big(q^{2\delta}+q^2+2+q^{-2}+q^{-2\delta}\Big)\frac{[\delta][\delta-1]}{[2]}\,,
\nn\\
\ldots
\label{gviaa}
\ee
In general
\be
g^{(\delta)}_{i+1} = \sum_{m=0}^{\delta-i} a^{(\delta)}_{i+m}\cdot\xi_{m,i+m}\,,
\ \ \ \ i=1,\ldots,\delta
\label{expang1}
\ee
and
\be
\boxed{
\xi_{m,n}:=
\frac{[n][2n-m-1]!}{[m]![2n-2m+1]!} \cdot
\Big([n+1-m]\cdot(q^{2n}+q^{-2n})+[2][n-m]\Big)\,.
}
\label{xif}
\ee
Expressions~\eqref{expang1} are another form of $C$-polynomials~\eqref{C-pol1}.

One can straightforward express $g_j$ through $a_j$ using~\eqref{C-pol1} and obtain another form of~\eqref{expang1}:
\be
g_{\delta-k+1}^{(\delta)}
= \sum\limits_{j=0}^{k}a_{\delta-k+j}^{(\delta)}\cdot \eta_{j,k}
=  \sum\limits_{j=0}^{k}a_{\delta-k+j}^{(\delta)}\cdot
\left\{\sum\limits_{\Delta\vdash j}\ \ (-)^{l(\Delta)} \!\!\!\!\!\!\!\!\!\!
\sum\limits_{\overset{\{i_l\}=
\{\delta-k,\dots,\delta-k+j-1\}}{i_l+\Delta_l\,\leq\,\delta-k+j}}\prod\limits_{l=1}^{j}\sigma_{i_l,\Delta_l}
\right\}
\label{coefG}
\ee
where at the r.h.s. there stands the sum over all Young diagrams $\Delta$ of the size $j$. For example:
\be
\eta_{0,k}=1\,, \nn \\
\eta_{1,k}=-\sigma_{\delta-k,1}\,, \nn \\
\eta_{2,k}=\sigma_{\delta-k,1}\,\sigma_{\delta-k+1,1}-\sigma_{\delta-k,2}\,\sigma_{\delta-k+1,0}\,.
\ee

\subsubsection{Examples}\label{examples}

We now list the implications of general formulas (\ref{fundAl}) and (\ref{expang1})
for particular small values of $\delta$.
These expressions can be simpler to deal with in practical implications.

\medskip

$\bullet$ In the case of the \textbf{defect $\delta=1$}
\be
F^{(1)}_{[1]}(1,q) = a^{(1)}_{0} +   a^{(1)}_1\cdot (q^2+q^{-2})
\ee
and
\be
g^{(1)}_2(q) = a^{(1)}_1 \cdot \xi_{0,1} = a^{(1)}_1\,.
\label{gdef1}
\ee
We see that $F^{(1)}_{[1]}$ and $g^{(1)}_2\sim G^{(1)}_{[2]}$ are not independent. They are related by the Alexander $C$-polynomial (see Section~\ref{Cpols}).

The obvious question is (see Section~\ref{integra}): Are {\it all} integer pairs $(a^{(1)}_1,a^{(1)}_{0})$ allowed for knots? In other words, can one get an arbitrary integer for $g^{(1)}_2$
and an arbitrary integer symmetric Laurent polynomial of degree $1$ in $q^{\pm 2}$
for $F^{(1)}_{[1]}$?



\medskip

$\bullet$ For the \textbf{defect $\delta=2$}
\be
F^{(2)}_{[1]}(1,q) = a^{(2)}_{0} +   a^{(2)}_1\cdot (q^2+q^{-2})+    a^{(2)}_2\cdot (q^4+q^{-4})
\ee
and
\be
g^{(2)}_2(q) =\sum_{m=0}^{1} a^{(2)}_{m+1}\xi_{m,m+1}= a^{(2)}_1 \cdot \xi_{0,1} + a^{(2)}_2\cdot \xi_{1,2}
= a^{(2)}_1  + a^{(2)}_2\cdot [2](q^3+q^{-3})\,, \nn \\
g^{(2)}_3(q) =\sum_{m=0}^{0} a^{(2)}_{m+2}\xi_{m,m+2}= a^{(2)}_2 \cdot \xi_{0,2} = a^{(2)}_2\,.
\label{gdef2}
\ee
This time we can again ask if all integer triples $(a^{(2)}_2,a^{(2)}_1,a^{(2)}_{0})$ are allowed (see Section~\ref{integra}).
From the very beginning we see that $g^{(2)}_2(q)$ has degree 2, not 1 (in $q^{\pm 2}$) but only two parameters  $a^{(2)}_1, a^{(2)}_2$, so that a generic integer polynomial of degree $2$ {\it cannot} appear in the role of $g^{(2)}_2(q)$. 

\medskip

$\bullet$ For the \textbf{defect $\delta=3$}
\be
F^{(3)}_{[1]}(1,q) = a^{(3)}_{0} +   a^{(3)}_1\cdot (q^2+q^{-2})+    a^{(3)}_2\cdot (q^4+q^{-4}) +    a^{(3)}_3\cdot (q^6+q^{-6})
\ee
and
\be
g^{(3)}_2(q) =\sum_{m=0}^{2} a^{(3)}_{m+1}\xi_{m,m+1}
= a^{(3)}_1  + a^{(3)}_2\cdot [2](q^3+q^{-3}) + a^{(3)}_3\cdot [3](q^6+1+q^{-6})\,,   \nn \\
g^{(3)}_3(q) =\sum_{m=0}^{1} a^{(3)}_{m+2}\xi_{m,m+2}= a^{(3)}_2 \cdot \xi_{0,2} + a^{(3)}_3 \cdot \xi_{1,3}
= a^{(3)}_2 +  a^{(3)}_3 \cdot [3](q^4+q^{-4})\,,\nn \\
g^{(3)}_4(q) =\sum_{m=0}^{0} a^{(3)}_{m+3}\xi_{m,m+3}= a^{(3)}_3 \cdot \xi_{0,3} = a^{(3)}_3\,.
\ee
Again, $g_2^{(3)}$ and $g_3^{(3)}$ are not arbitrary integer Laurent polynomials. In the next section, we will discuss whether all integer $a_j$ can be met in the symmetric Alexander DE.

%

\subsection{Intermediate summary}

We have proved that the defect $\delta$ defines the degree of the fundamental Alexander polynomial. Moreover, {\bf all the dependence on a knot ${\cal K}$ for all symmetric Alexander polynomials
is concentrated in $\delta$ parameters (numbers) $a^{(\delta)}_j$, $j=1,\ldots,\delta$}.
Note that  any given  $a^{(\delta)}_j$ enters only $F_{[1]}(1,q)$ and the first few
$G_{[i]}(1,q)$ with $i\leq j$.
This triangularity implies that the degree of $F_{[1]}$
defines the number of non-vanishing $G_{[i]}$.
Only $a^{(\delta)}_\delta$ appears in all $\delta$ DE coefficients $G_{[i]}(1,q)$,
which contribute to the Alexander polynomial
(note that the other $G_{[i]}(1,q)$ with $i>\delta+1$ need not vanish,
but they do not contribute to the DE at $A=1$).

Relations (\ref{expang1}), (\ref{coefG}) are somewhat sophisticated.
A useful way to represent them is through recurrence conditions (equations),
which are also known as $C$-polynomials \cite{Cpols} for the Alexander polynomials,
see Section \ref{Cpols} below.
Still, another question to address is if the same parameters $a^{(\delta)}_i$ are sufficient
to describe the reduced Alexander polynomials in all other representations.
We remind that the {\it reduced} Alexander polynomials are non-vanishing for
all representations.
The simplifying reduction condition, similar to (\ref{symAl}), remains true for all
single-hook representations, but it breaks violently for multiple hooks.
The last direction  will be addressed elsewhere, while one-hook representations will be considered in Section~\ref{otherreps}.
In the next section of the present paper
we will restrict to a kind of inverse question --
if everything about the Alexander polynomials is fully controlled by $a^{(\delta)}_i$,
then are they {\it free} parameters or somehow constrained as well?


%

\setcounter{equation}{0}
\section{Integrality
\label{integra}}

A remarkable property of knot polynomials is that all their coefficients are {\it integer}.
For the HOMFLY polynomials, this follows technically from the $R$-matrix formalism.
In general, this is the ground for cohomological descriptions in terms of various complexes,
like in  Khovanov-Rozansky approach \cite{khovanov2007virtual}.

In other words, all the coefficients $a_j$ in (\ref{fundAl}) are integers and
all $g_j$ in (\ref{defg}) are Laurent polynomials in $q$ with integer coefficients. From~\eqref{expang1}, it is not obvious why $g_j^{(\delta)}$ are integer Laurent polynomials in $q$, because of the monstrous coefficients $\xi_{m,n}$~\eqref{xif}. Still, there is a bit simpler formula~\eqref{coefG}. From this expansion, it is clear that $g_j^{(\delta)}$ are integer Laurent polynomials in $q$. Indeed, one rewrites $\sigma_{j,k}$ in the form
\be
\sigma_{j,k}
= \sum\limits_{i=0}^k(-1)^k q^{k(k-2i+1)-2i}\begin{bmatrix}
	j+k \\ i
\end{bmatrix}_{q^{-2}}\begin{bmatrix}
	j+k \\ j+i
\end{bmatrix}_{q^{2}},
\ee
so that  $\sigma_{j,k}$ is constructed with quantum binomial coefficients which are integer Laurent polynomials in $q$.

Now, there is a question whether one can get {\it any} integers as the values of $a$ and {\it any} Laurent polynomials $g$ with integer coefficients. In the following subsections we make a first attempt to understand the abundance of values of $a^{(\delta)}_i$
in the space of knots.
We look at the simplest possibility -- at the families of antiparallel descendants of 2-strand torus knots with the hope that they represent big enough sets for each particular defect.
The result turns out to be modest if not discouraging --
the values of $a^{(\delta)}_i$ appearing in these families look rather poor, at least for $\delta>0$.
Still, the exercise is interesting and it also provides a {\bf more accurate formulation for the defect-preservation conjecture of \cite{arxiv.2205.12238}} (see Section~\ref{defpre}),
	{\bf which makes it fully consistent with the defect-degree correspondence} studied in the present paper.




\subsection{ 2-strand torus knots}

The $(2m+1)_1$ torus knot (a closure of 2 strands with $2m+1$ crossings)
has maximal possible (for given number of intersections) defect $\delta=m-1$
and
\be
a^{(m-1)}_i = -{\rm floor}\left(\frac{m+1-i}{2}\right), \ \ \ \ i=0,\ldots,m-1\,.
\label{a2torus}
\ee
For example, for $9_1$ parameter $m=4$ and
\be
a^{(3)}_0=-{\rm floor}(5/2)=-2, \ \ \ \
a^{(3)}_1=-{\rm floor}(4/2)=-2, \ \ \ \
a^{(3)}_2=-{\rm floor}(3/2)=-1, \ \ \ \
a^{(3)}_3=-{\rm floor}(2/2)=-1\,.
\nn
\ee
Substituting (\ref{a2torus}) into (\ref{fundAl}) and (\ref{expang1}), we get
\be\label{torusGen}
F^{(2m+1)_1}_{[1]}(1,q) \ \stackrel{(\ref{fundAl})}{=}\ a^{(m-1)}_0 + \sum_{i=1}^{m-1} a^{(m-1)}_i(q^{2i}+q^{-2i})
\ \stackrel{(\ref{a2torus})}{=}\ -\frac{[m+1][m]}{[2]}\,,\nn \\
G^{(2m+1)_1}_{[2]}(1,q)
= -\frac{[m+2]!}{[3][4][m-2]!}\{q\}\,,
\nn \\
G^{(2m+1)_1}_{[3]}(1,q)
= -\frac{[m+3]!}{[4][5][6][m-3]!}\{q\}^2\,,
\nn \\
\ldots \nn \\
G^{(2m+1)_1}_{[i]}(1,q)  =  -\frac{[m+i]![i]!  }{[m-i]]![2i]!}\{q\}^{i-1}\,,
\nn \\
G^{(2m+1)_1}_{[m-j]}(1,q)  =  -\frac{[2m-j]![m-j]!  }{[j]![2m-2j]!}\{q\}^{m-j-1}\,,
\nn \\
\ldots\nn \\
G^{(2m+1)_1}_{[m-1]}(1,q)  \ \stackrel{(\ref{expang1})}{=}\
\Big(a^{(m-1)}_{m-2}+a^{(m-1)}_{m-1}\cdot[m-1](q^m+q^{-m})\Big)\prod_{i=2}^{m-1}\{q^i\}
\ \stackrel{(\ref{a2torus})}{=}\
-[2m-1][m-1]!\{q\}^{m-2}\,,
\nn \\
G^{(2m+1)_1}_{[m]}(1,q)  \ \stackrel{(\ref{expang1})}{=}\
a^{(m-1)}_{m-1}\cdot \prod_{i=2}^{m}\{q^i\}
\ \stackrel{(\ref{a2torus})}{=}\ \prod_{i=2}^{m}\{q^i\} = [m]!\{q\}^{m-1}\,.
\ee
All these DE coefficients are nicely factorized.

\subsection{Antiparallel descendants of torus knots}

The set (\ref{a2torus}) provides just one point per defect in $m=\delta+1$-dimensional
space of parameters $a^{m-1}_i$.
However, this may be not a big problem:
we can now use the defect-preserving antiparallel evolution \cite{2204.05977} in each intersection,
which gives rise to a $2m+1$-dimensional
family of antiparallel pretzels with the same defect $\delta=m-1$.
This dimension is more than enough,
the question is only if arbitrary integer vectors $a^{m-1}_i$ appear in this family.
Regarding $G_{[k]}$, as we have first seen in Subsection \ref{examples},
they are not generic integer symmetric Laurent polynomials of a given degree --
the degree is typically higher than the number of contributing parameters $a^{(\delta)}_i$.

According to \cite{mironov2015colored}, the HOMFLY polynomial for the odd antiparallel pretzel
of genus $2m$ is equal to
\be
\mathcal{H}_{[r]}^{(n_1,\ldots,n_{2m+1})} = \sum_{j=1}^{r+1}  \frac{d_{[r]}^{2m}}{d_{j}^{m-1/2}}
\prod_{a=1}^{2m+1} \left(\bar S \bar T^{2n_a-1} S\right)_{1,j}\,.
\label{arborpretz}
\ee
We restrict it to a symmetric representation $R=[r]$
and refer to Section 3 of \cite{arxiv.2205.12238} for detailed notation in this case.
We also do not put bars over $n_a$, what is usually done to distinguish {\it anti}parallel evolution,
since this is the only one which is matter for the formulas.
To further simplify them, we write $n_a$ in the labels and in the denotations of pretzel knots instead of $2n_a-1$.

For fundamental representation, we obtain:
\be\label{H1}
 \mathcal{H}_{[1]}^{(n_1,\ldots,n_{2m+1})} =\frac{\{Aq^{-1}\}}{\{q^2\}}\prod\limits_{a=1}^{2m+1}\frac{A^{2n_a-2}q^{-1}-q A^{2n_a}+q-q^{-1}}{\{A\}}+\frac{\{A q\}}{\{q^2\}}\prod\limits_{a=1}^{2m+1}\frac{-q A^{2n_a-2}+A^{2n_a}q^{-1}+q-q^{-1}}{\{A\}}\,.
\ee  
Taking limit $A\rightarrow 1$, we get that the degree of $q^{\pm 2}$ in $F_{[1]}$ is equal to $m$, and the corresponding coefficients of $F_{[1]}$ are given by formulas~\eqref{3-pretzel},~\eqref{aadef1} and~\eqref{aadefm-1}-\eqref{VEx}.

Strictly speaking, the defect is not fully preserved by antiparallel evolution.
For example, the trefoil $3_1$ is not only the parent-knot in defect-zero family,
but also a descendant of all other knots:  $3_1=(1,1,1)=(1,1,1,1,0)=(1,1,1,1,1,0,0) = \ldots$
As we will see (Section~\ref{defpre}), there are also other defect-zero points in higher-defect evolution families.
What is important for this paper, these dropdowns are easily captured by the fundamental Alexander --
its power also drops.

\subsection{Defect $\delta=m-1=0$ -- descendants of $3_1$
}\label{def0}

In this case there is just a single coefficient $a^{(0)}_0$ and all $g^{(0)}_j=0$.
Thus, the question is just if all integer $a_0^{(0)}$ are possible. The answer is positive, and an example is provided just by the family of twist knots
\cite{1306.3197}, which form a class of descendants of $3_1$:
\be
a^{(0)}_0\big({\rm twist}_n\big)=F_{[1]}^{{\rm twist}_n}(q,A=1) = n\,.
\ee
In particular,
\be
\ldots \nn \\
F_{[1]}^{7_2}(q,A=1) = -3\,, \nn \\
F_{[1]}^{5_2}(q,A=1) = -2\,, \nn \\
F_{[1]}^{3_1}(q,A=1) = -1\,, \nn \\
F_{[1]}^{{\rm unknot}}(q,A=1) = 0\,, \nn \\
F_{[1]}^{4_1}(q,A=1) = 1\,, \nn \\
F_{[1]}^{6_1}(q,A=1) = 2\,, \nn \\
F_{[1]}^{8_1}(q,A=1) = 3\,. \nn \\
\ldots
\ee
More examples are given by the larger  families of double braids \cite{Morozov_2016}
and 3-pretzels \cite{arxiv.2205.12238},
also obtained by the defect-preserving antiparallel evolution from the trefoil $3_1$ \cite{2204.05977}
and thus all having the defect zero:
\be\label{3-pretzel}
a^{(0)}_0\big({\rm pretzel}_{2l-1,2m-1,2n-1}\big)= lmn-(l-1)(m-1)(n-1)\,.
\label{3pretza}
\ee
They provide more delicate information about the abundancy/frequency of particular values
of $a^{(0)}_0$ in the population of knots -- which is a question at the next level of complexity.

\subsection{Defect $\delta=m-1=1$ -- descendants of $5_1$
	\label{desc51}}

Using (\ref{H1}) for defect $\delta=m-1=1$ (descendants of $5_1$) and the notation
\be
F_{[1]}^{(n_1,\ldots,n_5)}(1,q) = a_0^{(n_1,\ldots,n_5)} + a_1^{(n_1,\ldots,n_5)} (q^2+q^{-2})\,,
\ee
we get
\be
a_0^{(n_1,\ldots,n_5)}
= 2\sum_{1\leq a<b<c<d\leq 5}n_an_bn_cn_d  - 2\sum_{1\leq a<b<c\leq 5} n_an_bn_c
+ \sum_{1\leq a<b\leq 5} n_an_b  -  1\,,
\nn \\
a_1^{(n_1,\ldots,n_5)}
= -\sum_{1\leq a<b<c<d\leq 5}n_an_bn_cn_d  +\sum_{1\leq a<b<c\leq 5} n_an_bn_c
- \sum_{1\leq a<b\leq 5} n_an_b + \sum_{1\leq a\leq 5} n_a - 1\,.
\label{aadef1}
\ee
In particular,
\be
a_0^{(n_1,2,1,1,1)} = n_1-2, \ \ \ \   a_1^{(n_1,2,1,1,1)} = -2n_1\,,
\ee
what provides us with
a full-fledged 1-dimensional subspace in the 2-dimensional space
of integer $(a_0,a_1)$.
For our analysis of the Alexander polynomial for this family, equation (\ref{aadef1}) could be practical.
Still, it is rather difficult to extract conclusive statements.
Our preliminary impression is that the 2-dimensional set of $(a_0,a_1)$ for descendants of $5_1$
is severely restricted and does not cover the full integer lattice.
For example, for $a_1=\pm 2$ there are only odd $a_0$ in this family, moreover, they seem rather rare:
we have found $a_0=43,17,7,5,-1,-11$ for $a_1=-2$ and $a_0=-1,-3,-9$ for $a_1=2$
in the range $-10\leq n_1,\ldots, n_5\leq 10$.




\subsection{Defect $\delta=m-1>1$ -- descendants of $(2m+1)_1$
	\label{desc71}}

Using (\ref{H1}), we can calculate
\be
F_{[1]}^{(n_1,\ldots,n_{2m+1})}(1,q) = a_0^{(n_1,\ldots,n_{2m+1})}
+ \sum_{i=1}^{m-1} a_i^{(n_1,\ldots,n_{2m+1})} (q^{2i}+q^{-2i})
\ee
for any $m$.
It appears that the highest coefficient is always very simple:
\be
a_{m-1}^{(n_1,\ldots,n_{2m+1})} =  (-)^{m+1}\left(\prod_{a=1}^{2m+1} n_a - \prod_{i=1}^{2m+1} (n_a-1)\right)\,,
\label{aadefm-1}
\ee
For other coefficients
it is convenient to use the following parametrization:
\be\label{General_a}
a_{i}^{(n_1,\ldots,n_{2m+1})} = \sum_{k=1}^{2m} \ u_k^{(i)}\cdot
\left(\sum_{1\leq a_1<\ldots<a_k\leq 2m+1}  n_{a_1}\cdot\ldots \cdot n_{a_k}\right)\,.
\ee
Then
\be
(\ref{aadefm-1}): \ \ \ \  u_k^{(m-1)} = (-)^{k+1 }  \nn \\
u_k^{(m-2)} =  (-)^{k} (k-1-\delta_{k,2m})  \nn \\
u_k^{(m-3)} =  (-)^{k+1} \left(\frac{k^2-3k+4}{2} -\delta_{k,2m-1} +(2m-1)\delta_{k,2m})\right)  \nn \\
u_k^{(m-4)} =  (-)^{k} \left(\frac{k^3-6k^2+17k-12}{6}-\delta_{k,2m-2} -(2m-2)\delta_{k,2m-1}
-(2m^2-3m+2)
\delta_{k,2m}\right)  \nn \\
u_k^{(m-5)} =  (-)^{k+1} \left(\frac{k^4-10 k^3+47k^2-86k+72}{24}-\delta_{k,2m-3} -(2m-3)\delta_{k,2m-2}
\right.
-(2m^2-5m+4) \delta_{k,2m-1}
- \nn \\ \left.
-\frac{(2m-1)(2m^2-5m+6)}{3}\delta_{k,2m}\right)  \nn 
\ee
\vspace{-0.3cm}
\be
\ldots
\label{uexas}
\ee
$\bullet$ For example, (\ref{aadef1}) gives for $m=2$ (descendants of $5_1$):
\be
\begin{array}{c|cccccc }
	k& =&0&1&2&3&4  \\
	\hline
	u_k^{(1)} &=& -1&1&-1&1&-1  \\
	u_k^{(0)} &=& -1&0&1&-2&2 \\
\end{array}
\ee
$\bullet$ while for $m=3$ (descendants of $7_1$):
\be
\begin{array}{c|cccccccc}
	k& =&0&1&2&3&4&5&6 \\
	\hline
	u_k^{(2)} &=& -1&1&-1&1&-1&1&-1 \\
	u_k^{(1)} &=& -1&0&1&-2&3&-4&4 \\
	u_k^{(0)} &=& -2&1&-1&2&-4&6&-6 \\
\end{array}
\ee
$\bullet$ for $m=4$ (descendants of $9_1$):
\be
\begin{array}{c|cccccccccc}
	k& =&0&1&2&3&4&5&6&7&8 \\
	\hline
	u_k^{(3)} &=& -1&1&-1&1&-1&1&-1&1&-1 \\
	u_k^{(2)} &=& -1&0&1&-2&3&-4&5&-6&6 \\
	u_k^{(1)} &=& -2&1&-1&2&-4&7&-11&15&-15 \\
	u_k^{(0)} &=& -2&0&1&-2&4&-8&14&-20&20 \\
\end{array}
\ee
$\bullet$ for $m=5$ (descendants of $11_1$):
\be
\begin{array}{c|cccccccccccc}
	k& =&0&1&2&3&4&5&6&7&8&9&10 \\
	\hline
	u_k^{(4)} &=& -1&1&-1&1&-1&1&-1&1&-1&1&-1 \\
	u_k^{(3)} &=& -1&0&1&-2&3&-4&5&-6&7&-8&8 \\
	u_k^{(2)} &=& -2&1&-1&2&-4&7&-11&16&-22&28&-28 \\
	u_k^{(1)} &=& -2&0&1&-2&4&-8&15&-26&41&-56&56 \\
	u_k^{(0)} &=& -3&1&-1&2&-4&8&-16&30&-50&70&-70 \\
\end{array}
\ee
and so on.
The general case can be described as follows:
\be
u_k^{(m-j)} = (-)^{k+j}\left(U_j(k)-\sum_{i=0}^{j-2}   V_{j,i}(m)\delta_{k,2m-(j-2-i)}\right),
\ee
where the coefficients $U_j$ and $V_{j,i}$ are polynomials
of degree $j-1$ in $k$ and degree  $i$ in $m$ respectively,
which are defined recursively:
\be
U_{j+1}(k+1)-U_{j+1}(k) = U_j(k)\,,\nn\\
U_{j}(k=1) = \left\{\begin{array}{ccc} 0 &{\rm for \ even} & j \\ 1 & {\rm for \ odd} & j
\end{array}\right.
\ \ \ \  {\rm or} \ \ \ \ U_{j}(k=0) =(-)^{j+1} {\rm floor}\left(\frac{j+1}{2}\right),   \nn \\
V_{j+1,i}(m) = V_{j,i}\left(m-\frac{1}{2}\right), \nn \\
U_j(2m-1) -V_{j,j-3}(m) = U_j(2m)- V_{j,j-2}(m)\,,
\ee
where the forth constraint says that the  last two coefficients in front of $\delta_{k,2m-1}$
and $\delta_{k,2m}$ are the same.
In particular
\be\label{VEx}
V_{j,0}=1\, \nn \\
V_{j,1} = 2m-(j-2)\, \nn \\
V_{j,2} = 2m^2 - (2j-5)m + \frac{j^2-5j+8}{2}\,, \nn \\
\ldots
\ee
in accordance with the examples in (\ref{uexas}).

Expressions~\eqref{General_a} restrict integer sets $\{a_0,\dots,a_\delta\}$, so that for $\delta>0$ they are not arbitrary, at least for the family of antiparallel descendants of torus knots. In the next subsection we will provide a kind of probable explanation for sets of $a_j$ to be so restricted.


\subsection{Example of $9_1$}\label{9_1}

To emphasize the peculiarity of torus knots, we add an example of
\be
{\rm defect\ 3}:  & \text{knot } 9_1 &    a^{(3)}_0 = -2,\ \  a^{(3)}_1=-2,\ \  a^{(3)}_2=-1,\ \  a^{(3)}_3 = -1\,.
\ee
The point is that all the four non-vanishing combinations
are nicely factorized expressions~\eqref{torusGen}:
\be
F^{9_1}_{[1]}(1,q) = -\frac{[4][5]}{[2]}\,,\nn \\
G^{9_1}_{[2]}(1,q) = -\{q\}[5][6]\,, \nn \\
G^{9_1}_{[3]}(1,q) = -\{q\}^2[2][3][7]\,, \nn \\
G^{9_1}_{[4]}(1,q) = -\{q\}^3[2][3][4]\,.
\ee
 In this case, this factorization is a result of fine tuning of the coefficients $a^{(3)}_i$,
which are very special.
It can happen that the antiparallel descendants continue to carry some weakened traces
of this peculiarities and this is the reason why they do not describe the general situation
with a given defect.
In any case, the question, if the coefficients $a^{(\delta)}_i$ are all independent
or not, remains open.










\setcounter{equation}{0}
\section{Consequences for the defect-preservation conjecture}\label{defpre}

In this section, we discuss the defect drop downs and additional peculiarities to the standard defect-diagrams~\eqref{d=0}-\eqref{d=3}. The analysis is simpler to conduct by the use of the connection with the degree of the fundamental Alexander polynomial. Namely, the mentioned deviations occur if several highest coefficients $a_j^{(\delta)}$~\eqref{fundAl} vanish. In this case, there are two options:
\begin{enumerate}
	\item The defect drops down.
	\item The defect stays the same, but $F_{[1]}$ and some of $G_{[s]}$ turn out to additionally factorize.
\end{enumerate}
Below, we will consider concrete examples and discuss these properties in more detail. It turns out that some mentioned peculiarities and drop downs can be classified, so that such ones are not occasional.

\subsection{Descendants of $3_1$}\label{3_1Des}
In this case, just a $1_Z$-parametric family of twist knots within
the $3_Z$-dimensional set of trefoil descendants is sufficient to provide
any value of $a_0$ (see Subsection~\ref{def0}).
Still, triple pretzels provide additional curious examples.
As already mentioned in  \cite{2204.05977}, there are knots with the unit Alexander polynomial,
$a_0=0$, in this family, which are not unknots, for example\footnote{Recall that for pretzel knots $(2n_1-1,\dots,2n_{2m+1}-1)$, we denote $F$- and $G$-factors superscripts as $(n_1,\dots,n_{2m+1})$.}
\be
F_{[1]}^{(-1,3,4)}(A,q) = \{A\}A^3(A^2+1)(A^4+A^2+1)
\ee
and vanishes at $A=1$\footnote{Actually, for triple pretzels, all the knots have $F_{[1]}(A,q)$ independent of $q$
for arbitrary $A$ --
not only their Alexander limit   $F_{[1]}(1,q)$ at $A=1$.
However, this is not a general property of the defect-zero knots --
already in the next subsection \ref{desc51} we will mention the members of $5_1$ family,
where $F_{[1]}(A,q)$ are linear in $q^{\pm 2}$, as all the members of that family.
Still, the defect is $\delta=0$, because this $q$-dependence disappears at $A=1$.
Just a couple of non-trivial examples is provided by
\be
\delta^{(3,2,-1,-4,-5)}=0 & {\rm and} &
F_{[1]}^{(3,2,-1,-4,-5)} = 2 + \{A\}\Big(U_0(A)+U_1(A)(q^2+q^{-2})\Big)\,, \nn \\
\delta^{(3,2,-1,-3,-8)}=0 & {\rm and} &
F_{[1]}^{(3,2,-1,-3,-8)} = 2 + \{A\}\Big(V_0(A)+V_1(A)(q^2+q^{-2})\Big)
\ee
with somewhat complicated polynomials $U$ and $V$.

Then, there are defect-0 knots among the antiparallel descendants of $7_1$, with $F_{[1]}$ depending on the squares of $q^{\pm 2}$ at $A\neq 1$ and so on.
}. Actually, this knot is a point of a whole family $(-k,2k+1,2k+2)$. Moreover,
there are many other evolution points, where the Alexander polynomial turn to $1$.
Formally, we could associate to these cases a new sort of a diagram:
\be
\begin{picture}(300,100)(-40,-60)
\put(0,-20){
	\put(-88,20){\mbox{sometimes for}}
	\put(-90,0){\mbox{defect $\delta^{\cal K}=0$:}}
	\put(-100,0){
		\put(180,-20){\line(1,0){75}}
		\put(180,-10){\line(1,0){75}}
		\put(200,0){\line(1,0){55}}
		\put(210,10){\line(1,0){45}}
		\put(220,20){\line(1,0){35}}
		\put(230,30){\line(1,0){25}}
		\put(240,40){\line(1,0){15}}
		\put(250,50){\line(1,0){5}}
		\put(180,-20){\line(0,1){10}}
		\put(190,-20){\line(0,1){10}}
		\put(200,-20){\line(0,1){20}}
		\put(210,-20){\line(0,1){30}}
		\put(220,-20){\line(0,1){40}}
		\put(230,-20){\line(0,1){50}}
		\put(240,-20){\line(0,1){60}}
		\put(250,-20){\line(0,1){70}}
		\put(160,-35){\mbox{$s$}}
		\put(183,-35){\mbox{{\footnotesize $1$}}}
		\put(193,-35){\mbox{{\footnotesize $2$}}}
		\put(203,-35){\mbox{{\footnotesize $3$}}}
		\put(213,-35){\mbox{{\footnotesize $4$}}}
		\put(223,-35){\mbox{{\footnotesize $5$}}}
		\put(233,-35){\mbox{{\footnotesize $6$}}}
		\put(243,-35){\mbox{{\footnotesize $7$}}}
		\put(253,-35){\mbox{{\footnotesize $8$}}}
		%
		\put(-50,0){
			\put(320,-18){\mbox{{\footnotesize $1$}}}
			\put(320,-8){\mbox{{\footnotesize $2$}}}
			\put(320,2){\mbox{{\footnotesize $3$}}}
			\put(320,12){\mbox{{\footnotesize $4$}}}
			\put(320,22){\mbox{{\footnotesize $5$}}}
			\put(320,32){\mbox{{\footnotesize $6$}}}
			\put(320,42){\mbox{{\footnotesize $7$}}}
		}
		%
		%
}}
\label{degF1for3pret}
\end{picture}
\ee
In this way, one can actually reveal a more sophisticated structure in the defect
(like a \textit{Young diagram}, not just a single number).

\subsection{Descendants of $5_1$}\label{5_1Des}
With the help of (\ref{aadef1}),
one immediately observes  numerous zeroes of $a_1$ within the 5-fold antiparallel family --
does this signal about a deviation from the defect-preservation theorem?
Not quite, this rather calls for a more accurate formulation.
Antiparallel evolution can, of course, convert $5_1=(1,1,1,1,1)$ into $3_1=(1,1,1,1,0)$
and likewise the other knots from the defect-one family into those from defect-zero.
Thus, the defect-preservation means that the defect is not {\it increased}, but it can drop down
at particular values of evolution parameters.
This never happens when they all have the same signs, otherwise one should be carefull
with this kind of the defect drop-downs.

More interesting are the cases like $(3, 2, -1, -4, -5)$ and $(3, 2, -1, -3, -8)$
with $F_{[1]}(1,q)=2$ -- they have defect zero, but do not belong to the triple-pretzel family.
Unlike triple-pretzels, they have $q$-dependent $F_{[1]}(A,q)$, just $q$-dependence vanishes
at $A=1$.

Surprisingly or not, but we have not found any examples with the anomalous diagram
\be
\begin{picture}(300,75)(-40,-55)
\put(0,-20){
	\put(-88,20){\mbox{HYPOTETICAL:}}
	\put(-88,0){\mbox{sometimes for}}
	\put(-90,-20){\mbox{defect $\delta^{\cal K}=1$}}
	\put(-100,0){
		%
		\put(180,-20){\line(1,0){135}}
		\put(180,-10){\line(1,0){135}}
		\put(220,0){\line(1,0){95}}
		\put(240,10){\line(1,0){75}}
		\put(260,20){\line(1,0){55}}
		\put(280,30){\line(1,0){35}}
		\put(300,40){\line(1,0){15}}
		\put(180,-20){\line(0,1){10}}
		\put(190,-20){\line(0,1){10}}
		\put(200,-20){\line(0,1){10}}
		\put(210,-20){\line(0,1){10}}
		\put(220,-20){\line(0,1){20}}
		\put(230,-20){\line(0,1){20}}
		\put(240,-20){\line(0,1){30}}
		\put(250,-20){\line(0,1){30}}
		\put(260,-20){\line(0,1){40}}
		\put(270,-20){\line(0,1){40}}
		\put(280,-20){\line(0,1){50}}
		\put(290,-20){\line(0,1){50}}
		\put(300,-20){\line(0,1){60}}
		\put(310,-20){\line(0,1){60}}
		%
		\put(160,-35){\mbox{$s$}}
		\put(183,-35){\mbox{{\footnotesize $1$}}}
		\put(193,-35){\mbox{{\footnotesize $2$}}}
		\put(203,-35){\mbox{{\footnotesize $3$}}}
		\put(213,-35){\mbox{{\footnotesize $4$}}}
		\put(223,-35){\mbox{{\footnotesize $5$}}}
		\put(233,-35){\mbox{{\footnotesize $6$}}}
		\put(243,-35){\mbox{{\footnotesize $7$}}}
		\put(253,-35){\mbox{{\footnotesize $8$}}}
		\put(263,-35){\mbox{{\footnotesize $9$}}}
		\put(270,-35){\mbox{{\footnotesize $10$}}}
		\put(280,-35){\mbox{{\footnotesize $11$}}}
		\put(290,-35){\mbox{{\footnotesize $12$}}}
		\put(300,-35){\mbox{{\footnotesize $13$}}}
		\put(310,-35){\mbox{{\footnotesize $14$}}}
		\put(320,-18){\mbox{{\footnotesize $1$}}}
		\put(320,-8){\mbox{{\footnotesize $2$}}}
		\put(320,2){\mbox{{\footnotesize $3$}}}
		\put(320,12){\mbox{{\footnotesize $4$}}}
		\put(320,22){\mbox{{\footnotesize $5$}}}
		\put(320,32){\mbox{{\footnotesize $6$}}}
		%
		%
}}
\end{picture}
\ee
among the descendants of $5_1$.
The DE coefficients $F_{[1]}\sim \{A\}$ and vanishes for the Alexander polynomial
only for the knots of the type
$(-k,2k+1,2k+2,1,0)$ which have defect 0 and anomalous diagram
(\ref{degF1for3pret}), or for $(k,0,1,0,1)$ which are just unknots.
It is currently unclear if other kinds of anomalous diagrams, e.g.
\be
\begin{picture}(300,75)(-40,-50)
\put(0,-20){
	\put(-88,20){\mbox{HYPOTETICAL:}}
	\put(-88,0){\mbox{sometimes for}}
	\put(-90,-20){\mbox{defect $\delta^{\cal K}=1$}}
	\put(-100,0){
		\put(185,-16){\circle*{5}}
		\put(190,-20){\line(1,0){125}}
		\put(190,-10){\line(1,0){125}}
		\put(220,0){\line(1,0){95}}
		\put(240,10){\line(1,0){75}}
		\put(260,20){\line(1,0){55}}
		\put(280,30){\line(1,0){35}}
		\put(300,40){\line(1,0){15}}
		%
		\put(190,-20){\line(0,1){10}}
		\put(200,-20){\line(0,1){10}}
		\put(210,-20){\line(0,1){10}}
		\put(220,-20){\line(0,1){20}}
		\put(230,-20){\line(0,1){20}}
		\put(240,-20){\line(0,1){30}}
		\put(250,-20){\line(0,1){30}}
		\put(260,-20){\line(0,1){40}}
		\put(270,-20){\line(0,1){40}}
		\put(280,-20){\line(0,1){50}}
		\put(290,-20){\line(0,1){50}}
		\put(300,-20){\line(0,1){60}}
		\put(310,-20){\line(0,1){60}}
		%
		\put(160,-35){\mbox{$s$}}
		\put(183,-35){\mbox{{\footnotesize $1$}}}
		\put(193,-35){\mbox{{\footnotesize $2$}}}
		\put(203,-35){\mbox{{\footnotesize $3$}}}
		\put(213,-35){\mbox{{\footnotesize $4$}}}
		\put(223,-35){\mbox{{\footnotesize $5$}}}
		\put(233,-35){\mbox{{\footnotesize $6$}}}
		\put(243,-35){\mbox{{\footnotesize $7$}}}
		\put(253,-35){\mbox{{\footnotesize $8$}}}
		\put(263,-35){\mbox{{\footnotesize $9$}}}
		\put(270,-35){\mbox{{\footnotesize $10$}}}
		\put(280,-35){\mbox{{\footnotesize $11$}}}
		\put(290,-35){\mbox{{\footnotesize $12$}}}
		\put(300,-35){\mbox{{\footnotesize $13$}}}
		\put(310,-35){\mbox{{\footnotesize $14$}}}
		\put(320,-18){\mbox{{\footnotesize $1$}}}
		\put(320,-8){\mbox{{\footnotesize $2$}}}
		\put(320,2){\mbox{{\footnotesize $3$}}}
		\put(320,12){\mbox{{\footnotesize $4$}}}
		\put(320,22){\mbox{{\footnotesize $5$}}}
		\put(320,32){\mbox{{\footnotesize $6$}}}
		%
		%
}}
\end{picture}
\ee
also appear within the family generated by $5_1$.
Clearly, one needs to consider richer family of defect-one knots,
which is beyond the scope of the present paper.

\setcounter{equation}{0}
\section{Recurrence relations ($C$-polynomials) for the colored Alexanders
\label{Cpols}}

For a given knot, the HOMFLY polynomial for different representations are not
independent -- infinite set of relations between them is named
"quantum ${\cal A}$-polynomial", because in the quasiclassical limit
it reproduces the well-known topological invariant.
However, from the point of view of the knot moduli space,
the knot polynomials are in any way superficial,
and it makes more sense to look at the more fundamental variables.
As explained in \cite{artdiff}, the DE coefficients are better suited
for this purpose. Still, they are not free, and relations between  them
are named $C$-polynomials \cite{arxiv.2205.12238,Garoufalidis_2006,Cpols}
(the word "quantum" is omitted because classical topology does not seem to tell anything interesting for description of their quasiclassical limit).

Relations survive even in the case $A=1$, i.e. the $C$-polynomials
exist even for the Alexander polynomials.
However, in this case they can be studied exhaustively --
and actually this was done in Subsection \ref{exprforG}
of the present paper.
As we show there, at least for symmetric representations,
the Alexander $C$-polynomials --
relations between $G_{[r]}(1,q)$ with different $r$,
depend only on the {\it defect}, but not on the other details
of knot topology.
This is exactly what we want from the future ideal relations
between the colored HOMFLY polynomials -- to separate the dependence on the
representation from that one of the knot,
and in the case of the Alexander polynomial this is partly done by separation
the defect $\delta$ from the variables $a^{(\delta)}_j$ at given $\delta$.
It is yet unclear whether this completes the story,
because we did not yet manage to understand if $a^{(\delta)}_j$
are {\it free} integers or there are further relations between them.
Actually, we have shown that they are free for $\delta=0$,
and this is the case where we can claim that we know a
{\bf complete set of the Alexander $C$-polynomials}.
The situation with other $\delta$ remains to be clarified.

Another question is to find the relation between Alexander
and HOMFLY $C$-polynomials -- whether the former ones can be lifted
to the latter ones and whether the latter ones can be restricted to the former ones.
It is even unclear which set is bigger.
In fact, the theory of $C$-polynomials is just at its very beginning,
they are very hard to find, the known examples look ugly
and are very difficult to study.
Moreover, in the HOMFLY case they split into different classes,
say, for given $A$, for given $q$, for certain combinations of those.
In this sense, our result for the Alexander case is very encouraging --
it proves that this kind of problems can have an exhaustive
and elegant solution.
Therefore, we hope for a new interest and  progress with $C$-polynomials
in foreseeable future.

\setcounter{equation}{0}
\section{Other one-hook representations
\label{otherreps}}
In this section, we consider relations between coefficients of the DE for other one-hook representations $[r,1^L]$.

\subsection{Generalities}

As we have already mentioned near (\ref{symAl}),
this scaling relation is actually true not only for symmetric representations,
but for arbitrary 1-hook diagrams \cite{IMMM,Zhu_2013}:
\be
{\cal A}_{[r,1^{s-1}]}(q) = {\cal A}_{[1]}(q^{r+s-1})
\ \stackrel{(\ref{HfundDE})}{=}\ 1 - F_{[1]}(1,q^{r+s-1})\{q^{r+s-1}\}^2
= \nn \\
 \stackrel{(\ref{fundAl})}{=} \
 1 \ -\  [r+s-1]^2\{q \}^2\cdot
 \left( a^{(\delta)}_{0} + \sum_{j=1}^\delta a^{(\delta)}_j \cdot (q^{2j(r+s-1)}+q^{-2j(r+s-1)}\right).
\label{Alrs}
\ee
This allows us to impose restrictions on the DE coefficients in the case of
such representations.
The main problem here is that these representations are not rectangular,
there is a complication with multiplicities,
and not so much is known about the DE for them.

In fact, the matching is not quite trivial, because \cite{Morozov_2019}
\begin{itemize}
    \item[(i)] there are multiplicities, i.e.   $Q$ in the sum
\be
H_{R} = \sum_{Q\subset {R+X}}  F_{Q} Z^{Q }_{R}
\ee
is not in one-to-one correspondence with sub-diagrams of $R$,
\item[(ii)] since now the number of $Q$ exceeds that one of $R$
there is an ambiguity in  the coefficients $F_Q$
and
\item[(iii)]  the $Z$-factors do not vanish automatically at $A=1$
even for defect $\delta=0$.
\end{itemize}
See~\cite{morozov2020kntz,BM1} for further available details.

\subsection{Representation $R=[2,1]$ for the defect $\delta=0$}\label{[2,1],d=0}

For the simplest representation $[21]$:
\be
Z_{[2,1]}^{[0]} = 1, \ \ \ \
Z_{[2,1]}^{[1]} = \frac{[3]\{A\}^2+[3]^2\{Aq^2\}\{A/q^2\}}{[2]^2}\,, \nn \\
Z_{[2,1]}^{[2]} = \frac{[3]}{[2]}\{Aq^3\}\{Aq^2\}\{A\}\{A/q^2\}, \ \ \ \
Z_{[2,1]}^{[1,1]} = \frac{[3]}{[2]} \{Aq^2\}\{A\}\{A/q^2\}\{A/q^3\}\,, \nn \\
Z_{[2,1]}^{[2,1] } = \{Aq^3\}\{Aq^2\}\{Aq\}\{A/q\}\{A/q^2\}\{A/q^3\}, \ \ \ \
Z_{[2,1]}^{X_2 } = -[3]^2\{q\}^4\{Aq^2\}\{A/q^2\}\,,
\ee
and only the $Z$-factors in the second line vanish at $\{A\}=0$
(this matters when the defect is zero and all $F_Q$ are not singular at $A=1$).
Thus, for the Alexander polynomial we have:
\be
{\cal A}_{[2,1]}^{(\delta=0)}
= 1 - [3]^2\{q\}^2 F_{[1]}^{(\delta=0)}(1,q)
- [3]^2[2]^2\{q\}^6 \Big(F_{[2,1] }^{(\delta=0)}(1,q) -  F_{X_2 }^{(\delta=0)}(1,q)\Big)\,.
\label{21def0}
\ee
The very different $Z$-factors $Z_{[2,1]}^{[2,1] }$ and $Z_{[2,1]}^{X_2 }$
almost coincide at $A=1$ (just differ in sign), the ambiguity in the $F$-coefficients
reduces to coinciding shifts of $F_{[2,1]}(1,q)$ and $F_{X_2}(1,q)$
(and certain change of $F_{[2,2]}(1,q)$, which is beyond our consideration here) --
in the result, what is well defined is just the difference $F_{[2,1]}(1,q) -F_{X_2}(1,q)$.

Since from (\ref{def0a}) we know that for defect $\delta=0$
the DE coefficient $F_{[1]}^{(\delta=0)}(1,q)$
is actually independent of $q$, comparison with (\ref{Alrs}) implies that
\be
F_{[2,1] }^{(\delta=0)}(1,q) = F_{X_2 }^{(\delta=0)}(1,q)\,.
\label{F2-FX2}
\ee
Indeed, for defect-zero twist knots ${\rm Tw}_m$, the DE coefficients
$F_{[2,1]}$ and $F_{X_2}$ are the sums of elements
of respectively the fifth and the sixth lines of the KNTZ triangular matrix $B_{[2,1]}^{m+1}$,
see (13) of \cite{Morozov_2019},
\be
B_{[2,1]}=\left(\begin{array}{cccccc}
1 & 0 & 0 & 0 & 0 & 0 \\ \\
-A^2& A^2& 0& 0& 0& 0 \\ \\
\frac{A^4}{q^2}& -\frac{[2]A^4}{q^3}& \frac{A^4}{q^4}& 0& 0& 0 \\ \\
A^4q^2& -[2]q^3A^4& 0& A^4q^4& 0& 0 \\ \\
-A^6& [3]A^6& -\frac{[3]A^6}{[2]q}& -\frac{[3]A^6q}{[2]}& A^6& 0\\ \\
-A^6& [3]A^6& \frac{(A^2-q^6)A^4}{q^2(q^4-1)}& -\frac{(A^2q^6 - 1)A^4}{q^4=1}& 0& A^4
\end{array}\right)\,,
\ee
and it is easy to check that for all integer $m$ the difference
\be
F_{[2,1]}^{{\rm Tw}_m}-F_{X_2}^{{\rm Tw}_m} \sim \{A\}\,,
\ee
i.e. it vanishes at $A=1$.
The proportionality coefficient is quite a complicated polynomial of $A^{\pm 1}$ and $q^{\pm 1}$,
thus, the statement does not look trivial.

We can now generalize it in two directions: to other 1-hook representations (see Subsections~\ref{[r,1],d=0} and~\ref{[r,1^L],d=0})
and to non-vanishing defects (see Subsection~\ref{[2,1],d>0}).

\subsection{Representation $R=[2,1]$ for the defect $\delta\geq 1$}\label{[2,1],d>0}
For non-vanishing defect, some DE coefficients $F^Q_R$ become singular at $A=1$,
and there are more terms of the DE contributing to the Alexander polynomial.
This can be illustrated with the example of two-strand torus knots,
which have maximal defect per intersection.

In the case of $R=[2,1]$, we expect that for $\delta\geq 1$
the coefficients $F_{[2]}$ and $F_{[1,1]}(q)=F_{[2]}(q^{-1})$
are proportional to $\{A\}^{-1}$, so that there are two more contributions to
the generalization of (\ref{21def0}):
\be
{\cal A}_{[2,1]}^{(\delta>0)}
= 1 - [3]^2\{q\}^2 F_{[1]}^{(\delta>0)}(1,q)
-[3]^2[2]\{q\}^3\left(G_{[2]}^{(\delta>0)}(1,q)-G_{[2]}^{(\delta>0)}(1,q^{-1})\right)
- \nn \\
- [3]^2[2]^2\{q\}^6 \Big(F_{[2,1] }^{(\delta>0)}(1,q) -  F_{X_2 }^{(\delta>0)}(1,q)\Big)\,.
\label{21def>0}
\ee
This time, (\ref{Alrs}) means that
\be
F_{[1]}^{(\delta>0)}(1,q^3) = F_{[1]}^{(\delta>0)}(1,q)
+[2]\{q\}\left(G_{[2]}^{(\delta>0)}(1,q)-G_{[2]}^{(\delta>0)}(1,q^{-1})\right)
+[2]^2\{q\}^4 \Big(F_{[2,1] }^{(\delta>0)}(1,q) -  F_{X_2 }^{(\delta>0)}(1,q)\Big)
\ \ \Longrightarrow \nn \\
\sum_{j=1}^\delta a_j^{(\delta)}\left(q^{6j}+q^{-6j} - q^{2j}- q^{-2j}\right)
=[2]\{q\}\left(G_{[2]}^{(\delta>0)}(1,q)-G_{[2]}^{(\delta>0)}(1,q^{-1})\right)
+[2]^2\{q\}^4 \Big(F_{[2,1] }^{(\delta>0)}(1,q) -  F_{X_2 }^{(\delta>0)}(1,q)\Big)\,.
\nn
\ee
In particular, for $\delta=1$ we get from the identity
$q^6+q^{-6}-q^2-q^{-2} = (q^2-q^{-2})^2\left(2 +\{q\}^2\right)$ :
\be
G_{[2]}^{(\delta=1)}(1,q) = a_1^{(1)}(q^2-q^{-2})\,, \nn \\
F_{[2,1] }^{(\delta=1)}(1,q) -  F_{X_2 }^{(\delta=1)}(1,q) =  a_1^{(1)}\,.
\label{21def1}
\ee
Indeed, for the simplest defect-1 torus knot $5_1$ we have  $a_1^{5_1}=-1$, and
(\ref{21def1}) is satisfied.
The first statement in (\ref{21def1}) is the same as (\ref{gdef1}).

Likewise, for $\delta=2$ we get
\be
a_1^{(2)}(q^6+q^{-6}-q^2-q^{-2}) + a_2^{(2)}(q^{12}+q^{-12}-q^4-q^{-4}) = \nn \\
=[2]\{q\}\left(G_{[2]}^{(\delta>0)}(1,q)-G_{[2]}^{(\delta>0)}(1,q^{-1})\right)
+[2]^2\{q\}^4 \Big(F_{[2,1] }^{(\delta>0)}(1,q) -  F_{X_2 }^{(\delta>0)}(1,q)\Big)\,.
\label{21def2}
\ee
This relation {\it per se} is not sufficient to find both terms in the second line,
but we can substitute (\ref{gdef2}) for
\be
G_{[2]}(1,q) = \Big(a_1^{(2)} + a_2^{(2)}\cdot[2](q^3+q^{-3})\Big)\{q^2\}
\ee
to get
\be
F_{[2,1] }^{7_2}(1,q) -  F_{X_2 }^{7_2}(1,q) =
a^{(2)}_1 + a^{(2)}_2\cdot\left(q^6+2q^4+3q^2+2+3q^{-2}+2q^{-4}+q^{-6}\right)\,.
\ee
Indeed, for defect-2 torus knot $7_1$ we have $a_1^{7_1}=a_2^{7_1}=-1$, and
\be
G_{[2]}^{7_2}(1,q) = -[5](q^2-q^{-2})\,,\nn \\
F_{[2,1] }^{7_2}(1,q) -  F_{X_2 }^{7_2)}(1,q) =  -[3][5]\,.
\ee

\subsection{Representations $R=[r,1]$ for the defect $\delta=0$}\label{[r,1],d=0}

For the defect $\delta=0$ and any $r$, the only
non-vanishing at $A=1$ is the quadruple $Z^{[0]}_{[r,1]}=1$, $Z^{[1]}_{[r,1]}$,
$Z^{[2,1]}_{[r,1]}$ and $Z^{X_2}_{[r,1]}$, so that
\be
{\cal A}_{[r,1]}^{(\delta=0)}
= 1 - [r+1]^2\{q\}^2 F_{[1]}^{(\delta=0)}(1,q)
- [2][r+1]^2[r][r-1]\{q\}^6 \Big(F_{[2,1] }^{(\delta=0)}(1,q) -  F_{X_2}^{(\delta=0)}(1,q)\Big)\,.
\ee
Comparison with (\ref{Alrs}) gives:
\be
F_{[2,1] }^{(\delta=0)}(1,q) =  F_{X_2}^{(\delta=0)}(1,q)\,,
\ee
which is just the same  (\ref{F2-FX2}) for all $r$.


\subsection{Representations $R=[r,1^{s-1}]$ for the defect $\delta=0$}\label{[r,1^L],d=0}
First, let us obtain the DE for the Alexander polynomial for $R=[r,1^{s-1}]$ and the defect $\delta=0$. For this representation:
\begin{equation}
    \mathcal{H}^\mathcal{K}_{[r,1^{s-1}]}(q,A=q^{L+1})=\mathcal{H}^\mathcal{K}_{[r-1]}(q,A=q^{s})\;\Longrightarrow \mathcal{H}^\mathcal{K}_{[r,1^{s-1}]}(q,A)-\mathcal{H}^\mathcal{K}_{[r-1]}(q,A)\sim \{A q^{-s}\}\,.
\end{equation}
Looking at lots of examples, we induce that 
\begin{equation}\label{A0}
	\mathcal{A}^\mathcal{K}_{[r,1^{s-1}]}=\mathcal{A}^\mathcal{K}_{[r-1]}+G_{[1]}(1,q)\{q^{-s}\}\{q^{2r+s-2}\}=1-F_{[1]}(1,q)\{q^{r-1}\}^2+G_{[1]}(1,q)\{q^{-s}\}\{q^{2r+s-2}\}\,.
\end{equation}
For example, we have:
\begin{itemize}
	\item for $3_1$: $G_{[1]}(1,q)=-1$
	\item for $4_1$: $G_{[1]}(1,q)=1$
	\item for $5_2$: $G_{[1]}(1,q)=-2$
\end{itemize}
Note that result~\eqref{A0} are in a straightforward correspondence with the results of Subsections~\ref{[2,1],d=0} and~\ref{[r,1],d=0}.

Comparing~\eqref{A0} and~\eqref{Alrs}, we obtain
\begin{equation}
    G_{[1]}(1,q)=F_{[1]}(1,q)=a_0^{(\delta)}\,.
\end{equation}

\setcounter{equation}{0}
\section{Conclusion
\label{conc}}

The purpose of this paper is to study the  implications
of the differential expansion (\ref{DE}) for the special case
of the symmetrically-colored Alexander polynomials ($A=1$),
which satisfy (\ref{symAl}).
In other words, we studied  the combined implications
of (\ref{DE}) and (\ref{symAl}).

The main result is the proof of
the conjecture \cite{Kononov_2015} that the defect $\delta$,
which was originally defined as characteristic of factorization depth
of the coefficients of differential expansion
for the HOMFLY polynomials in symmetric representations,
can be alternatively related to the degree $\delta+1$ of the fundamental Alexander polynomial (see Subsection~\ref{DofAl}).
We have provided explicit formula for all other DE coefficients at $A=1$
in terms of the fundamental Alexander polynomial and explain that they are {\it not} free (see Subsection~\ref{exprforG}).

We address the question, if the {\it fundamental}  Alexander coefficients  {\it are} free,
and as an example demonstrate that for $\delta=0$ they {\it are} (see Subsection~\ref{def0}).
We use for this purpose the family of antiparallel descendants of the trefoil.
Our results in this case can be interpreted as the
{\bf full separation of representation and knot variables},
i.e. a {\bf complete set of $C$-polynomials} (see Section~\ref{Cpols}).

Looking at different descendants of the trefoil we also observe the defect peculiarities, which make the statement of the defect-preserving theorem more precise (see Section~\ref{defpre}). 

Similar calculations for other defects (not only $\delta=0$),
other representations (not only symmetric)
and other knot polynomials (not only Alexanders)
remain to be done.
We describe just the simplest steps in these directions (see Subsections~\ref{desc51}-\ref{9_1} and Section~\ref{otherreps}),
which prove that they are indeed interesting and doable. 

\section*{Acknowledgements}

We would like to thank Nikita Tselousov for useful discussions. Our work was supported by the Russian Science Foundation (Grant No.21-12-00400).

\printbibliography

@article{BM1,
    author = "Bishler, L. and Morozov, A.",
    title = "{Perspectives of differential expansion}",
    eprint = "2006.01190",
    archivePrefix = "arXiv",
    primaryClass = "hep-th",
    doi = "",
    journal = "Phys. Lett. B",
    volume = "808",
    pages = "135639",
    year = "2020"
}

@article{IMMM,
    author = "Itoyama, H. and Mironov, A. and Morozov, A. and Morozov, An.",
    title = "{HOMFLY and superpolynomials for figure eight knot in all symmetric and antisymmetric representations}",
    eprint = "1203.5978",
    archivePrefix = "arXiv",
    primaryClass = "hep-th",
    reportNumber = "FIAN-TD-04-12, ITEP-TH-14-12, OCU-PHYS-364",
    doi = "",
    journal = "JHEP",
    volume = "07",
    pages = "131",
    year = "2012"
}

@article{garoufalidis2005analytic,
  title={An analytic version of the Melvin-Morton-Rozansky conjecture},
  author={Garoufalidis, S. and Le, Thang TQ},
  eprint = {math/0503641},
  archivePrefix = {arXiv},
    primaryClass = {math.GT},
  year={2005}
}

@article{garoufalidis2011asymptotics,
  title={Asymptotics of the colored Jones function of a knot},
  author={Garoufalidis, S. and L{\^e}, Thang T Q},
  journal={Geometry \& Topology},
  volume={15},
  number={4},
  pages={2135--2180},
  year={2011},
  publisher={Mathematical Sciences Publishers},
  eprint = {math/0508100},
  archivePrefix = {arXiv},
    primaryClass = {math.GT}
}

@article{UFN3,
    author = "Morozov, A.",
    title = "{Integrability and matrix models}",
    eprint = "hep-th/9303139",
    archivePrefix = "arXiv",
    doi = "",
    journal = "Phys. Usp.",
    volume = "37",
    pages = "1-55",
    year = "1994"
}

@article{Cpols,
    author = "Mironov, A. and Morozov, A.",
    title = "{Algebra of quantum $ \mathcal{C} $-polynomials}",
    eprint = "2009.11641",
    archivePrefix = "arXiv",
    primaryClass = "hep-th",
    reportNumber = "FIAN/TD-13/20; IITP/TH-13/20; ITEP/TH-18/20; MIPT/TH-12/20",
    doi = "",
    journal = "JHEP",
    volume = "02",
    pages = "142",
    year = "2021"
}

@article{CS,
    author = "Chern, S.-S. and Simons, J.",
    title = "{Characteristic forms and geometric invariants}",
    doi = "",
    journal = "Annals Math.",
    volume = "99",
    pages = "48--69",
    year = "1974"
}

@article{Garoufalidis_2006,
	doi = {},
	year = 2006,
	month = {oct},
  
	publisher = {Mathematical Sciences Publishers},
  
	volume = {6},
  
	number = {4},
  
	pages = {1623--1653},
  
	author = {S. Garoufalidis and X. Sun},
  
	title = {The C-polynomial of a knot},
	journal = {Algebraic Geometric Topology},
	eprint = {math/0504305},
    archivePrefix = {arXiv},
    primaryClass = {math.GT}
}

@article{Witten,
  title={Quantum field theory and the Jones polynomial},
  author={Witten, E.},
  journal={Communications in Mathematical Physics},
  volume={121},
  number={3},
  pages={351--399},
  year={1989},
  publisher={Springer},
  doi={}
}

@article{alexander1928topological,
  title={Topological invariants of knots and links},
  author={Alexander, J.W.},
  journal={Transactions of the American Mathematical Society},
  volume={30},
  number={2},
  pages={275--306},
  year={1928},
  publisher={JSTOR}
}

@article{leech1970computational,
  title={Computational problems in abstract algebra},
  author={Algebraic Properties, J.H. Conway},
  year={1970},
  journal={John Leech (ed.), Proc.Conf.Oxford, Pergamon Press, Oxford-New York},
  pages={329-358},
  publisher={Citeseer}
}

@article{jones1983invent,
  title={Invent. Math.},
  author={Jones, V.F.R.},
  journal={Index for subfactors},
  volume={72},
  pages={1--25},
  year={1983}
}

@article{kauffman1987state,
  title={State models and the Jones polynomial},
  author={Kauffman, L.H.},
  journal={Topology},
  volume={26},
  number={3},
  pages={395--407},
  year={1987},
  publisher={Elsevier}
}

@article{freyd1985new,
  title={A new polynomial invariant of knots and links},
  author={Freyd, P. and Yetter, D. and Hoste, J. and Lickorish, WB R. and Millett, K. and Ocneanu, A.},
  journal={Bulletin (new series) of the American mathematical society},
  volume={12},
  number={2},
  pages={239--246},
  year={1985},
  publisher={American Mathematical Society}
}

@article{przytycki1987kobe,
  title={Kobe J. Math.},
  author={Przytycki, J.H. and Traczyk, K.P.},
  journal={Invariants of links of Conway type},
  volume={4},
  pages={115--139},
  year={1987},
	eprint = {1610.06679},
    archivePrefix = {arXiv},
    primaryClass = {math.GT}
}

@article{morozov2016there,
  title={Are there p-adic knot invariants?},
  author={Morozov, A.},
  journal={Theoretical and Mathematical Physics},
  volume={187},
  number={1},
  pages={447--454},
  year={2016},
  publisher={Springer},
	eprint = {1509.04928},
    archivePrefix = {arXiv},
    primaryClass = {hep-th}
}

@article{reshetikhin1990ribbon,
  title={Ribbon graphs and their invaraints derived from quantum groups},
  author={Reshetikhin, N.Yu. and Turaev, V.G.},
  journal={Communications in Mathematical Physics},
  volume={127},
  number={1},
  pages={1--26},
  year={1990},
  publisher={Springer}
}

@article{guadagnini1990clausthal,
  title={Clausthal 1989, Procs. 307-317},
  author={Guadagnini, E. and Martellini, M. and Mintchev, M.},
  journal={Phys. Lett. B},
  volume={235},
  pages={275},
  year={1990}
}

@article{turaev1992state,
  title={State sum invariants of 3-manifolds and quantum 6j-symbols},
  author={Turaev, V.G. and Viro, O.Ya.},
  journal={Topology},
  volume={31},
  number={4},
  pages={865--902},
  year={1992},
  publisher={Pergamon}
}

@article{morozov2010chern,
  title={Chern--Simons theory in the temporal gauge and knot invariants through the universal quantum R-matrix},
  author={Morozov, A. and Smirnov, A.},
  journal={Nuclear Physics B},
  volume={835},
  number={3},
  pages={284--313},
  year={2010},
  publisher={Elsevier},
	eprint = {1001.2003},
    archivePrefix = {arXiv},
    primaryClass = {hep-th}
}

@incollection{smirnov2012notes,
  title={Notes on Chern-Simons theory in the temporal gauge},
  author={Smirnov, A.},
  booktitle={The Most Unexpected at LHC and the Status of High Energy Frontier},
  pages={489--498},
  year={2012},
  publisher={World Scientific},
	eprint = {0910.5011},
    archivePrefix = {arXiv},
    primaryClass = {hep-th}
}

@incollection{mironov2013character,
  title={Character expansion for HOMFLY polynomials I: Integrability and difference equations},
  author={Mironov, A. and Morozov, A. and Morozov, And.},
  booktitle={Strings, gauge fields, and the geometry behind: the legacy of Maximilian Kreuzer},
  pages={101--118},
  year={2013},
  publisher={World Scientific},
	eprint = {1112.5754},
    archivePrefix = {arXiv},
    primaryClass = {hep-th}
}

@article{anokhina2013colored,
  title={Colored HOMFLY polynomials as multiple sums over paths or standard Young tableaux},
  author={Anokhina, A. and Mironov, A. and Morozov, A. and others},
  journal={Advances in High Energy Physics},
  volume={2013},
  year={2013},
  publisher={Hindawi},
	eprint = {1304.1486},
    archivePrefix = {arXiv},
    primaryClass = {hep-th}
}

@article{nawata2017colored,
  title={Colored HOMFLY-PT polynomials that distinguish mutant knots},
  author={Nawata, S. and Ramadevi, P. and Singh, V. K.},
  journal={Journal of Knot Theory and Its Ramifications},
  volume={26},
  number={14},
  pages={1750096},
  year={2017},
  publisher={World Scientific},
	eprint = {1504.00364},
    archivePrefix = {arXiv},
    primaryClass = {math.GT}
}

@article{mironov2015colored,
  title={Colored HOMFLY polynomials of knots presented as double fat diagrams},
  author={Mironov, A. and Morozov, A. and Morozov, And. and Ramadevi, P. and Singh, V. K.},
  journal={Journal of High Energy Physics},
  volume={2015},
  number={7},
  pages={1--70},
  year={2015},
  publisher={Springer},
	eprint = {1504.00371},
    archivePrefix = {arXiv},
    primaryClass = {hep-th}
}

@article{mironov2015towards,
  title={Towards effective topological field theory for knots},
  author={Mironov, A. and Morozov, A.},
  journal={Nuclear Physics B},
  volume={899},
  pages={395--413},
  year={2015},
  publisher={Elsevier},
	eprint = {1506.00339},
    archivePrefix = {arXiv},
    primaryClass = {hep-th}
}

@article{khovanov2000categorification,
  title={A categorification of the Jones polynomial},
  author={Khovanov, M.},
  journal={Duke Mathematical Journal},
  volume={101},
  number={3},
  pages={359--426},
  year={2000},
  publisher={Duke University Press},
	eprint = {math/9908171},
    archivePrefix = {arXiv},
    primaryClass = {math.QA}
}

@article{bar2002khovanov,
  title={On Khovanov’s categorification of the Jones polynomial},
  author={Bar-Natan, D.},
  journal={Algebraic \& Geometric Topology},
  volume={2},
  number={1},
  pages={337--370},
  year={2002},
  publisher={Mathematical Sciences Publishers},
	eprint = {math/0201043},
    archivePrefix = {arXiv},
    primaryClass = {math.QA}
}

@article{khovanov2004sl,
  title={sl (3) link homology},
  author={Khovanov, M.},
  journal={Algebraic \& Geometric Topology},
  volume={4},
  number={2},
  pages={1045--1081},
  year={2004},
  publisher={Mathematical Sciences Publishers},
	eprint = {math/0304375},
    archivePrefix = {arXiv},
    primaryClass = {math.QA}
}

@article{khovanov2010categorifications,
  title={Categorifications from planar diagrammatics},
  author={Khovanov, M.},
  journal={Japanese Journal of Mathematics},
  volume={5},
  number={2},
  pages={153--181},
  year={2010},
  publisher={Springer},
	eprint = {1008.5084},
    archivePrefix = {arXiv},
    primaryClass = {math.QA}
}

@article{khovanov2007virtual,
  title={Virtual crossings, convolutions and a categorification of the SO (2N) Kauffman polynomial},
  author={Khovanov, M. and Rozansky, L.},
  journal={arXiv preprint math/0701333},
  year={2007},
	eprint = {math/0701333},
    archivePrefix = {arXiv},
    primaryClass = {math.QA}
}

@article{dolotin2014introduction,
  title={Introduction to Khovanov homologies. III. A new and simple tensor-algebra construction of Khovanov--Rozansky invariants},
  author={Dolotin, V. and Morozov, A.},
  journal={Nuclear Physics B},
  volume={878},
  pages={12--81},
  year={2014},
  publisher={Elsevier},
	eprint = {1308.5759},
    archivePrefix = {arXiv},
    primaryClass = {hep-th}
}

@article{arxiv.2205.12238,
  doi = {},
  
  url = {},
  
  author = {Morozov, A. and Tselousov, N.},
  
  title = {Differential Expansion for antiparallel triple pretzels: the way the factorization is deformed},
  
  publisher = {arXiv},
  
  year = {2022},
  
  copyright = {arXiv.org perpetual, non-exclusive license},
	eprint = {2205.12238},
    archivePrefix = {arXiv},
    primaryClass = {hep-th}
}

@article{Morozov_2018,
	doi = {},
  
	url = {},
  
	year = 2018,
	month = {apr},
  
	publisher = {World Scientific Pub Co Pte Lt},
  
	volume = {33},
  
	number = {12},
  
	pages = {1850062},
  
	author = {A. Morozov},
  
	title = {Factorization of differential expansion for non-rectangular representations},
  
	journal = {Modern Physics Letters A},
	eprint = {1612.00422},
    archivePrefix = {arXiv},
    primaryClass = {hep-th}
}

@article{Morozov_2019,
	doi = {},
  
	url = {},
  
	year = 2019,
	month = {jun},
  
	publisher = {Elsevier {BV}
},
  
	volume = {793},
  
	pages = {464--468},
  
	author = {A. Morozov},
  
	title = {Extension of {KNTZ} trick to non-rectangular representations},
  
	journal = {Physics Letters B},
  publisher={Springer},
	eprint = {1903.00259},
    archivePrefix = {arXiv},
    primaryClass = {hep-th}
}

@article{morozov2020kntz,
  title={The KNTZ trick from arborescent calculus and the structure of the differential expansion},
  author={Morozov, A.},
  journal={Theoretical and Mathematical Physics},
  volume={204},
  number={2},
  pages={993--1019},
  year={2020},
  publisher={Springer},
	eprint = {2001.10254},
    archivePrefix = {arXiv},
    primaryClass = {hep-th}
}

@article{Kononov_2016,
	doi = {},
  
	url = {},
  
	year = 2016,
	month = {nov},
  
	publisher = {World Scientific Pub Co Pte Lt},
  
	volume = {31},
  
	number = {38},
  
	pages = {1650223},
  
	author = {Ya. Kononov and A. Morozov},
  
	title = {On rectangular {HOMFLY} for twist knots},
  
	journal = {Modern Physics Letters A},
	eprint = {1610.04778},
    archivePrefix = {arXiv},
    primaryClass = {hep-th}
}

@article{Morozov_2016,
	doi = {},
  
	url = {},
  
	year = 2016,
	month = {sep},
  
	publisher = {Springer Science and Business Media {LLC}
},
  
	volume = {2016},
  
	number = {9},
  
	author = {A. Morozov},
  
	title = {Factorization of differential expansion for antiparallel double-braid knots},
  
	journal = {Journal of High Energy Physics},
	eprint = {1606.06015},
    archivePrefix = {arXiv},
    primaryClass = {hep-th}
}

@article{Kameyama_2020,
	doi = {},
  
	url = {},
  
	year = 2020,
	month = {jul},
  
	publisher = {Springer Science and Business Media {LLC}
},
  
	volume = {110},
  
	number = {10},
  
	pages = {2573--2583},
  
	author = {M. Kameyama and S. Nawata and R. Tao and H. Derrick Zhang},
  
	title = {Cyclotomic expansions of {HOMFLY}-{PT} colored by rectangular Young diagrams},
  
	journal = {Letters in Mathematical Physics},
	eprint = {1902.02275},
    archivePrefix = {arXiv},
    primaryClass = {math.GT}
}

@article{arxiv.2110.03616,
  doi = {},
  
  url = {},
  
  author = {Q. Chen, K. Liu, Sh. Zhu},
  
  title = {Cyclotomic expansions for the colored HOMFLY-PT invariants of double twist knots},
  
  publisher = {arXiv},
  
  year = {2021},
  
  copyright = {Creative Commons Attribution 4.0 International},
	eprint = {2110.03616},
    archivePrefix = {arXiv},
    primaryClass = {math.GT}
}

@article{arxiv.1512.07906,
  doi = {},
  
  url = {},
  
  author = {Q. Chen},
  
  title = {Cyclotomic expansion and volume conjecture for superpolynomials of colored HOMFLY-PT homology and colored Kauffman homology},
  
  publisher = {arXiv},
  
  year = {2015},
  
  copyright = {arXiv.org perpetual, non-exclusive license},
	eprint = {1512.07906},
    archivePrefix = {arXiv},
    primaryClass = {math.QA}
}

@article{arxiv.2101.08243,
  doi = {},
  
  url = {},
  
  author = {Beliakova, A. and Gorsky, E.},
  
  title = {Cyclotomic expansions for $\mathfrak{gl}_N$ knot invariants via interpolation Macdonald polynomials},
  
  publisher = {arXiv},
  
  year = {2021},
  
  copyright = {Creative Commons Attribution 4.0 International},
	eprint = {2101.08243},
    archivePrefix = {arXiv},
    primaryClass = {math.RT}
}

@article{arxiv.1908.04415,
  doi = {},
  
  url = {},
  
  author = {Berest, Yu. and Gallagher, J. and Samuelson, P.},
  
  title = {Cyclotomic Expansion of Generalized Jones Polynomials},
  
  publisher = {arXiv},
  
  year = {2019},
  
  copyright = {arXiv.org perpetual, non-exclusive license},
	eprint = {1908.04415},
    archivePrefix = {arXiv},
    primaryClass = {math.QA}
}

@article{Kononov_2015,
	doi = {},
  
	url = {},
  
	year = 2015,
	month = {jun},
  
	publisher = {Pleiades Publishing Ltd},
  
	volume = {101},
  
	number = {12},
  
	pages = {831--834},
  
	author = {Ya. Kononov and A. Morozov},
  
	title = {On the defect and stability of differential expansion},
  
	journal = {{JETP} Letters},
	eprint = {1504.07146},
    archivePrefix = {arXiv},
    primaryClass = {hep-th}
}

@article{2204.05977,
      title={Evolution properties of the knot's defect}, 
      author={A. Morozov and N. Tselousov},
      year={2022},
      eprint={2204.05977},
      archivePrefix={arXiv},
      primaryClass={hep-th}
}

@article{superintegrarev,
  doi = {},
  
  url = {},
  
  author = {Mironov, A. and Morozov, A.},
  
  title = {Superintegrability summary},
  
  publisher = {arXiv},
  
  year = {2022},
  
  copyright = {arXiv.org perpetual, non-exclusive license},
	eprint = {2201.12917},
    archivePrefix = {arXiv},
    primaryClass = {hep-th}
}

@article{artdiff,
	doi = {},
  
	url = {},
  
	year = 2014,
	month = {may},
  
	publisher = {Springer Science and Business Media {LLC}
},
  
	volume = {179},
  
	number = {2},
  
	pages = {509--542},
  
	author = {S. B. Arthamonov and A. D. Mironov and A. Yu. Morozov},
  
	title = {Differential hierarchy and additional grading of knot polynomials},
  
	journal = {Theoretical and Mathematical Physics},
	eprint = {1306.5682},
    archivePrefix = {arXiv},
    primaryClass = {hep-th}
}

@article{MIRONOV_2010,
	doi = {},
  
	url = {},
  
	year = 2010,
	month = {jun},
  
	publisher = {World Scientific Pub Co Pte Lt},
  
	volume = {25},
  
	number = {16},
  
	pages = {3173--3207},
  
	author = {A. Mironov and A. Morozov and Sh. Shakirov},
  
	title = {{Conformal} {blocks} {as} {Dotsenko}{\textendash}{Fateev} {integral} {discriminants}
},
  
	journal = {International Journal of Modern Physics A},
	eprint = {1001.0563},
    archivePrefix = {arXiv},
    primaryClass = {hep-th}
}

@article{arxiv.1012.2636,
  doi = {},
  
  url = {},
  
  author = {Liu, K. and Peng, P.},
  
  title = {New Structure of Knot Invariants},
  
  publisher = {arXiv},
  
  year = {2010},
  
  copyright = {arXiv.org perpetual, non-exclusive license},
	eprint = {1012.2636},
    archivePrefix = {arXiv},
    primaryClass = {math.GT}
}

@inproceedings{1306.3197,
	doi = {},
  
	url = {},
  
	year = 2013,
	publisher = {{AIP}
},
  
	author = {A. Mironov and A. Morozov and And. Morozov},
  
	title = {Evolution method and {\textquotedblleft}differential hierarchy{\textquotedblright} of colored knot polynomials},
  
	booktitle = {{AIP} Conference Proceedings},
	eprint = {1306.3197},
    archivePrefix = {arXiv},
    primaryClass = {hep-th}
}

@article{Zhu_2013,
	doi = {},
  
	url = {},
  
	year = 2013,
	month = {oct},
  
	publisher = {Springer Science and Business Media {LLC}
},
  
	volume = {2013},
  
	number = {10},
  
	author = {Sh. Zhu},
  
	title = {Colored {HOMFLY} polynomials via skein theory},
  
	journal = {Journal of High Energy Physics},
	eprint = {1206.5886},
    archivePrefix = {arXiv},
    primaryClass = {math.GT}
}

@article{Mironov3_2016,
	doi = {},
  
	url = {},
  
	year = 2016,
	month = {feb},
  
	publisher = {Springer Science and Business Media {LLC}
},
  
	volume = {2016},
  
	number = {2},
  
	author = {A. Mironov and R. Mkrtchyan and A. Morozov},
  
	title = {On universal knot polynomials},
  
	journal = {Journal of High Energy Physics},
	eprint = {1510.05884},
    archivePrefix = {arXiv},
    primaryClass = {hep-th}
}

@article{Morozov_2020,
	doi = {},
  
	url = {},
  
	year = 2020,
	month = {dec},
  
	publisher = {Springer Science and Business Media {LLC}
},
  
	volume = {80},
  
	number = {12},
  
	author = {A. Morozov and N. Tselousov},
  
	title = {Are Maxwell knots integrable?},
  
	journal = {The European Physical Journal C},
	eprint = {2010.02165},
    archivePrefix = {arXiv},
    primaryClass = {hep-th}
}

@article{arxiv.2205.05650,
  doi = {},
  
  url = {},
  
  author = {Bishler, L. and Mironov, A. and Morozov, And.},
  
  title = {Invariants of knots and links at roots of unity},
  
  publisher = {arXiv},
  
  year = {2022},
  
  copyright = {Creative Commons Attribution 4.0 International},
	eprint = {2205.05650},
    archivePrefix = {arXiv},
    primaryClass = {hep-th}
}

@article{arxiv.2207.08242,
  doi = {},
  
  url = {},
  
  author = {Mironov, A. and Morozov, A.},
  
  title = {Superintegrability as the hidden origin of Nekrasov calculus},
  
  publisher = {arXiv},
  
  year = {2022},
  
  copyright = {arXiv.org perpetual, non-exclusive license},
	eprint = {2207.08242},
    archivePrefix = {arXiv},
    primaryClass = {hep-th}
}

@article{Dunin_Barkowski_2013,
	doi = {},
  
	url = {},
  
	year = 2013,
	month = {mar},
  
	publisher = {Springer Science and Business Media {LLC}
},
  
	volume = {2013},
  
	number = {3},
  
	author = {P. Dunin-Barkowski and A. Mironov and A. Morozov and A. Sleptsov and A. Smirnov},
  
	title = {Superpolynomials for torus knots from evolution induced by cut-and-join operators},
  
	journal = {Journal of High Energy Physics},
	eprint = {1106.4305},
    archivePrefix = {arXiv},
    primaryClass = {hep-th}
}

\end{document}